\documentclass[prd,twocolumn]{revtex4}
\usepackage{amsmath,amssymb,graphicx,bm,enumerate,color,slashed}

\newcommand{\be}{\begin{equation}}
\newcommand{\ee}{\end{equation}}
\newcommand{\ba}{\begin{eqnarray}}
\newcommand{\ea}{\end{eqnarray}}

\newcommand{\bk}{{\bf k}}

\begin{document}

\title{
The Magneto-Sono-Luminescence 
 and its signatures\\ in photon and dilepton production in heavy ion collisions  }

\author{ 
G\"ok\c ce Ba\c sar$^1$, Dmitri E. Kharzeev$^{1,2}$ and Edward V. Shuryak$^1$}

\affiliation{$^1$Department of Physics and Astronomy, \\ Stony Brook University,\\
Stony Brook, New York 11794, USA}

\affiliation{$^2$Department of Physics, \\ Brookhaven National Laboratory,\\
Upton, New York 11973, USA}

\date{\today}

\begin{abstract} 
The matter produced in the early stages of heavy ion collisions consists mostly of gluons, and is penetrated by  coherent magnetic field produced by spectator nucleons. 
The fluctuations of gluonic matter in an external magnetic field couple to real and virtual photons through virtual quark loops. We study the resulting contributions to photon and dilepton production that stem from the fluctuations of the stress tensor $T_{\mu\nu}$ in the background of a coherent magnetic field $\vec{B}$. Our study extends significantly the earlier work by 
two of us and Skokov \cite{Basar:2012bp}, in which only the fluctuations of the $trace$ of the stress tensor  $T_{\mu\mu}$ were considered (the coupling of $T_{\mu\mu}$ to electromagnetic fields is governed by the scale anomaly). 
In the present paper we derive more general relations
using the Operator Product Expansion (OPE). We also extend the previous study to the case of dileptons which offers the possibility to discriminate between various production mechanisms. Among the phenomena that we study are Magneto-Sono-Luminescence (MSL, the interaction of magnetic field $\vec{B}(x,t)$ with the sound perturbations of the stress tensor $\delta T_{\mu\nu}(x,t)$) and 
Magneto-Thermo-Luminescence (MTL,  the interaction of  $\vec{B}(x,t)$ with smooth 
average $<T_{\mu\nu}>$). We calculate the rates of these process and find that they can dominate the photon and dilepton production at early stage of heavy ion collisions. We also point out the characteristic signatures of MSL and MTL that can be used to establish their presence and to diagnose the produced matter.

 \end{abstract}

\maketitle
\section{Introduction}
\subsection{Overview}

Dileptons and photons are ``penetrating probes" originally proposed as the signature of QGP formation  \cite{Shuryak:1978ij}; unlike hadrons, they are not produced from the
final freeze-out surface, but from the bulk of the matter, throughout the entire history of the collision.
In QCD matter, only quarks possess electric charges and can produce dileptons and photons -- however these quarks 
may not only be real but also virtual, in the form of quantum loops in gluonic matter. In this paper we will focus on the latter, quantum contributions to the 
photon and dilepton production that originate from the presence of magnetic field created by spectator nucleons. 
 \vskip0.3cm
 Since the coherent magnetic field exists mostly at
 the early stage of the collision \cite{Kharzeev:2007jp,Skokov:2009qp}, 
 we will not discuss the later hadronic stage of the collision and various hadronic  mechanisms of photon/dilepton production. 
As explained  in \cite{Shuryak:1978ij,Rapp:1999zw}, the early stages contribute noticeably to dilepton production only in a certain window of invariant mass $M$: 
  \be m_\phi \approx 1 \, {\rm {\rm GeV}} < M <   m_\psi \approx 3 \, {\rm {\rm GeV}}  \label{IMD};\ee
  these are the {\em intermediate mass} (IM)  dileptons.
This range of $M$ is limited from below by the region in which hadronic processes dominate, and from above by  both the charmonium decay background and the Drell-Yan processes (prompt partonic production). 
\vskip0.3cm
In the case of photons, the transverse momentum spectrum at $p_\perp \sim$ few {\rm GeV}  is affected by
early thermal radiation with a high initial temperature, but is still ``contaminated" by the late-stage emission.  While the
photon transverse momenta can be boosted (``blue-shifted") by collective flow of matter, the dilepton mass $M$ is a Lorentz invariant quantity 
and cannot be affected by the collective flow. 
\vskip0.3cm
 Experimentally,  the detection of the IM dileptons  suffers from a background 
 arising from the leptonic charm 
decays. At CERN SPS  this issue was  finally resolved  by the NA60 experiment few years ago,
which has determined that most of the IM dileptons are in fact ``prompt", emitted from thermal
medium and not from charm decays.
At RHIC these issues still remain to be solved. PHENIX hopes to do so using new vertex detectors
and STAR is expected to use its ability to detect events with $e-\mu$ charm decays. 
At LHC the charm background is perhaps overwhelming, and will have to be removed.
\vskip0.3cm
Theoretically,  the production of photons and dileptons is tied to the presence of quarks, and is thus sensitive to the {\em quark chemical equilibration}.
 The initial stages of the  high energy collisions are believed to be dominated by gluons. 
 Perturbative arguments \cite{Shuryak:1992bt} show that chemical equilibration via quark-antiquark pair production
is relatively slow and should be delayed relative to thermal equilibration of the glue. 
This idea led to a scenario in which the quark/antiquark sector in early stage matter (so-called
``hot glue") is suppressed by some fugacity $\xi<1$. If so, 
  the basic process of the dilepton production
 \be q+\bar{q} \rightarrow \gamma^* \rightarrow l_++l_- \ee
  is expected to be suppressed quadratically, $\sim \xi^2$.
While the strongly coupled nature of QGP, which is related to early hydrodynamics, may invalidate the perturbative  arguments,
one is still motivated to search for processes that involve gluons and do not require the existence of quark and antiquarks in the medium.
\vskip0.3cm
   Lacking real quarks, one can think of higher order processes with virtual quark loops as an
intermediary between the glue and    the  electromagnetic signals. 
In a recent paper \cite{Chiu:2012ij} such
processes have been considered,  e.g. with three gluons converting into a dilepton: 
\be ggg \rightarrow ({\rm quark \, \, loop}) \rightarrow \gamma^* \rightarrow l_++l_-. \ee
Furthermore, the processes with the number of gluons $n>3$  -- permitted by the
 global conservation laws such as P and C parity -- are equally important,
 because in the ``glasma" picture used by
these authors, the gauge field is $O(1/\alpha_s) $,  
which compensates the powers of $\alpha_s$ in the loop. 
A detailed quantitative assessment of the rates of such processes would be of great interest.
\vskip0.3cm
Another development is the BKS approach 
\cite{Basar:2012bp} introducing  a photon production mechanism  due to the coupling to the trace of the stress tensor and the
ambient QED magnetic field 
induced by the scale anomaly (also a virtual quark loop):
\be T^\mu_\mu + \vec{B}  \rightarrow  ({\rm quark \, \, loop})   \rightarrow \gamma. \ee
Due to the scale anomaly, the trace of the stress tensor in the chiral limit is given by the scalar gluon operator. 
Therefore the initial state here can be considered as a scalar gluon pair $\sigma \equiv gg$ and a virtual photon (from $\vec B$), 
so the vertex is of the $\sigma \gamma \gamma$ type. Since magnetic field in heavy ion collisions is on the average directed 
perpendicular to the reaction plane \cite{Kharzeev:2007jp}, the photons are produced mainly in-plane \cite{Basar:2012bp}. For the discussion of the effects of axial anomaly in magnetic field on photon and dilepton production, see 
\cite{AA,Tuchin-m,Fukushima,Yee:2013qma,Mamo:2013jda,Kharzeev:2013wra,Yin:2013kya}
and a recent review 
\cite{Kharzeev:2013ffa}. Note also that the scale anomaly was argued to play an important role in the gluon distributions at small $x$ \cite{Kharzeev:1999vh}, and in the presence of magnetic field could lead to the photon production even before the hydrodynamical description becomes valid.

\vskip0.3cm
   The PHENIX collaboration at RHIC \cite{Adare:2011zr} and then the ALICE   collaboration at LHC \cite{Lohner:2012ct}  recently
   discovered an unexpectedly large second azimuthal coefficient
$v_2=<\cos(2\phi)>$ of the produced photons. This feature is  known as
  ``large photon elliptic flow puzzle", because the conventional photon sources
have a difficulty explaining these observations. 
 Since the BKS process predicts \cite{Basar:2012bp} a photon production rate which depends on the direction of the magnetic field,
it  generates a strong angular anisotropy for the emitted photons, and provides a possible explanation for this puzzle.
\vskip0.3cm
In this paper we follow the BKS ideas and extend them in several directions as follows:\\
(i) We include the full stress tensor $T_{\mu\nu}$ and not just its trace $T_{\mu\mu}$ that is suppressed for 
the case of near-conformal strongly coupled QGP and classical glue.
\\  
(ii) We implement the space-time dependence of the magnetic field into the kinematics of the process and
show that its short lifetime is directly related to the characteristic invariant masses and energies of the produced photons. \\
(iii) Last but not least, we will not restrict ourselves to
 on-shell photons $\gamma$. Virtual (positive-mass) photons $\gamma^*$ are
observable via the dilepton channel, and they add to the overall momentum $\vec q$ extra observables:
the dilepton mass and polarization. These observables are very valuable for the separation of various production mechanisms. 
\vskip0.3cm
Since we discuss different processes and kinematic domains, we 
feel that it is useful to introduce some new terminology.
 While the calculations presented below are based on a single effective
action, we will distinguish two types of processes by their kinematics. We will call the interaction of the ambient electromagnetic field and the 
``average'' matter stress tensor $<T_{\mu\nu}>$, producing photons ( real or virtual),   {\em Magneto-Thermo-Luminescence}, MTL for short. 
By ``average" we mean that the value of the stress tensor is averaged over the fireball and is nearly constant, with negligible
momentum harmonics $p\sim 1/R$. Individual events, however, are known to also possess fluctuations  of the  matter stress tensor $\delta T_{\mu\nu}$,
with complicated spatial distribution and thus non-negligible momenta. Although these fluctuations include both longitudinal and transverse 
modes, in a somewhat a loose way we will refer to all of them as ``sounds".   
We will thus call the interaction of the ambient electromagnetic field and the fluctuations  of the  matter stress tensor that produces photons and dileptons 
{\em Magneto-Sono-Luminescence}, MSL. 
\vskip0.3cm

The theoretical task is divided into two steps. The first step is the derivation of a local effective action from 
 a quark-loop-induced non-local action that couples two photons to two gluons, the $gg\gamma\gamma$ vertex.
 We first combine
the gluons  into three colorless combinations and then describe the coupling to photons by a local effective  action.
This can be done in two ways: using ``generalized vector dominance" in scalar and tensor hadronic channels, or
 using the Operator Product Expansion (OPE).
The latter is based on the approximation in which the photon momenta are considered to be large as compared to that of the gluons, so that the photon-gluon vertex is effectively local. 
Needless to say, all steps must be both QED and QCD gauge invariant. 
The second step is the derivation of the dilepton production rate using this effective Lagrangian. 
\vskip0.3cm

From the point of view of phenomenology, the main proposal of this paper is development of new diagnostic tools.
Neither the collective
magnetic field in the collisions, nor perturbations of  
the stress tensor are not studied
in a detailed quantitative manner so far. (We will briefly review what is known about those in the  subsections \ref{sec_magnetic} and \ref{sec_sounds}, respectively.)  
The MTS and  MSL phenomenona may relate them to  photons or dileptons which are directly observable.
\vskip0.3cm

The structure of the paper is as follows. In Section \ref{sec_kin}, we explain the basic kinematics of the processes we consider and present the outline of the calculation.
Sections \ref{sec_MTL_hadr} and \ref{sec_Pi} are devoted to the derivations of two alternative effective Lagrangians that couple the gluonic stress tensor to
two photons, the magnetic field and the produced (virtual) photon. In Section \ref{sec_MTL_hadr}, 
we follow a hadronic approach and use the tensor meson dominance to evaluate the $gg\gamma\gamma$ coupling. This section concludes with an order-of-magnitude comparison of the MTL rate to the conventional quark annihilation process. 
The core of the paper is section \ref{sec_Pi}, in which we use the Operator Product Expansion
(OPE) methods to derive the effective Lagrangian. In Section \ref{sec_MTL}, we use the OPE result to calculate the MTL rate and compare the OPE prediction to the prediction of the hadronic 
approach. In Section \ref{sec_MSL} we focus on the MSL process and discuss the stress tensor correlators. Our results are summarized in the concluding Section \ref{sec_conc}.

\subsection{Coherent magnetic field in heavy ion collisions} \label{sec_magnetic}

Coherent fields of the heavy ions are proportional to their charge $Z$ and also to Lorentz factor of the beam $\gamma$. These large factors
provide several orders of magnitude enhancement. 
Furthermore, in the middle point between two ions the electric fields are opposite and tend to cancel out, while magnetic fields are parallel and add up.
Thus matter produced in the collision is subject to strong magnetic fields \cite{Kharzeev:2007jp}. The field vanishes for central collisions $b=0$
and has well defined oriention -- normal to the beam and impact parameter: those features should help to identify the effects induced by it.
\vskip0.3cm
We further
 assume that the magnetic field is directed in the transverse direction $\mu=3$ (normal to impact parameter -- direction 2 --
and the beam  -- direction 1).   We  use the gauge in which $A_1=x_2 B_3=-i{\partial \over \partial k_2} B_3(k) $, so that
\be \tilde A^B_\nu = \delta_{\nu 1} \tilde B_{3,2}(q)   \ee
where for brevity we use notations in which the index after a comma denotes a partial derivative.
We will parameterize x-dependence of the magnetic field by a Gaussian, and $t$-dependence by a function (\ref{eqn_skokov}).
\vskip0.3cm
According to calculations \cite{Kharzeev:2007jp,Skokov:2009qp,Bzdak:2011yy,Tuchin:2013apa,Tuchin:2013ie,Gursoy:2014aka}
,  the magnetic field strength for RHIC energies  reaches $e B\sim 0.2 \, {\rm {\rm GeV}}^2= 10\ m_{\pi}^2$  at $b=10 \, $fm, where it has a maximum. This is only a few times less than that of the gluon fields in the ``glasma" $g G\sim Q_s^2 \sim 1-2 \, {\rm {\rm GeV}}^2$.
  Due to the same Lorentz factor, it has a decay time $t_D\sim1/\sqrt{s}$; note however that the combination of Faraday \cite{Tuchin:2013apa,Tuchin:2013ie,McLerran:2013hla} and Hall \cite{Gursoy:2014aka} effects significantly delays the decay of magnetic field in the (electrically conducting) quark-gluon plasma. In addition to the coherent field $B_3$ at the center of the collision, there are also fluctuations of $B_{2,3}$ due to the event-by-event fluctuations of the positions of the nucleons inside the source nuclei \cite{Bzdak:2011yy,Deng:2012pc}.  The sampling over events leads to the following time dependence of the magnetic field at the center of the collision:
\ba
B_{3}(t)={B^0_{3}\over 1+(t/t_B)^2} \label{eqn_skokov}
\ea
  The duration parameter is fitted to be \be t_B\approx 0.15\, fm \ee for RHIC full collision energy, which is roughly
the Lorentz contracted nuclear diameter. The Fourier transform of this time dependence is exponential $\sim exp(-\omega t_B)$. 
\vskip0.3cm
Due to the finite electric conductivity of the quark-gluon plasma, the magnetic field  
can be partially trapped by the QGP as a result of Faraday induction \cite{Tuchin:2013apa,Tuchin:2013ie,McLerran:2013hla}; a recent analysis \cite{Gursoy:2014aka} including the collective expansion of the fluid and the resulting Hall effect confirms the existence of this phenomenon and points out its signatures in directed flow of charged hadrons. Nevertheless, the strongest
magnetic field is achieved at early times as discussed above.
\vskip0.3cm

The energy scale conjugated to $t_B$ is about a {\rm GeV}, so one can indeed produce dileptons with
energies and masses in the IM dilepton window. The transverse momenta associated with the $B$ field
are however quite small, of the order of the inverse nuclear size $1/R\sim 30\, {\rm {\rm MeV}}$, as they are associated with the distance to the ``spectator" nucleons
creating this magnetic field. Therefore the transverse momenta of the produced dileptons 
in fact originate mainly not from magnetic field  $B$ but from the phonons.
\vskip0.3cm

Let us discuss further the spatial and time dependencies of magnetic field and the parameterizations that we use to describe them. 
 We assume for simplicity (neglecting fluctuations) that $B$ is in the transverse plane and normal to the impact parameter, thus it has direction 3 in our notation.
  We further assume that it has a Gaussian profile in the transverse plane
\be eB_3 =(eB) {exp(- x_2^2/R_2^2 - x_3^2/R_3^2) \over (1+x_0^2/t_B^2) (1+x_1^2/t_B^2)}
 \ee
 Following tradition, we combine the electron charge with the field strength; this will allow us to compare the strength of electromagnetic and strong forces; 
 $(eB)$ is the field magnitude, and 3 is an index indicating the direction of magnetic field.
In principle there are two different radii in  transverse directions 2 and 3, as we consider non-central collisions.
\vskip0.3cm
 We can now make a Fourier transform
\be e\tilde{B}_3= \pi^3 (eB) (R_2 R_3 t_B^2) 
\ee
$$\times \exp\left(- {p_2^2 R_2^2 \over 4} - {p_3^2 R_3^2 \over 4} -|p_0| t_B-|p_1| t_B \right)  ;$$
for the gauge potential $A_1=x_2 B_3$ using $x_2=-i \partial/\partial p_2$ we get
\be \tilde{A}_1( p) = {\pi^3 \over 2} (eB) p_2 R_2^3 R_3 t_B^2 \ee
$$\times \exp\left(- {p_2^2 R_2^2 \over 4} - {p_3^2 R_3^2 \over 4} -|p_0| t_B-|p_1| t_B \right)  $$
In the propagator of two magnetic potentials, we prefer to read backwards and absorb the 4-dimensional
volume into the definition, so that it becomes  a product of the Fourier transforms (dimension -6)
\be  <A^B_\mu A^B_{\mu '}> d^4 x \rightarrow | \tilde A^B_\mu(q) \tilde A^B_{\mu '}(q)| 
\ee

Let us now comment on the normalization of the yield. The Fourier transforms $\tilde A$ are space integrals
and are thus proportional to the 4-volume $V_4$ of the system. Since the rates include the square of $\tilde A$, the volume squared appears. After the 
integrals over momenta $q$ are performed, one gets another $V_4^{-1}$ factor, which restores the expected dependence of the total yield, $\sim  V_4$ .

\section{Kinematics and the outline of the calculation}
\label{sec_kin}
\subsection{Kinematics}
Before we embark on the rather involved calculation of the photon/dilepton rates, let us define the kinematics of the processes we will consider, and
discuss some qualitative features of the problem.
We  define the 4-momenta of leptons as $l^\mu_+$ and  $l^\mu_-$; the momentum conservation yields
\be 
l^\mu_+ + l^\mu_-= q^\mu=k^\mu+p^\mu,
\ee
where momenta $k$ and $p$ correspond to Fourier harmonics of the stress tensor and the magnetic field, respectively.
The stress tensor and the magnetic field are two components that carry out two different tasks in the kinematics of the production process. The magnetic field comes as a short burst
in time: therefore its frequency $p^0\sim 1/t_B$ is large and thus it contributes most of the energy of the produced dileptons $q^0\approx p^0$. However,
the magnetic field is near-homogeneous in space, and thus its 3-momentum   ${\bf p}$ is very small.
Therefore the dilepton momentum comes mostly from those of the sound harmonics, ${\bf q}\approx {\bf p}$. 
\vskip0.3cm

This ``separation of responsibilities" greatly simplifies the calculation. As we will see below,
the dilepton yield is proportional to complicated kinematical expressions times 
the space-time integral of the product of  magnetic field and  the matter energy density correlators.
We can approximate this integral by a factorized form  
$$ \int d^4x <B B>< \epsilon \epsilon>  $$
$$ \approx \left(\int dt <B B>\right) \left(\int d^3x < \epsilon \epsilon>\right)
$$ 
\be =
 \left(\int {dp^0 \over 2\pi} \tilde B^2(p^0) \right) \left(\int {d^3k \over (2\pi)^3} \tilde\epsilon^2(k)\right), 
\ee
where in the first bracket we have (mostly) the time Fourier component, and in the second - mostly the space one.
The energy and momentum conservation then translates these components into dilepton observables.
\vskip0.3cm

To preview our subsequent discussion, let us indicate that
the MSL dilepton  yield appears proportional to the product of two factors: 
\be {dN\over d^4q} \sim \exp\left(-2 q^0 t_B- {c_s |{\bf q}| \over T_i}\right) \ee 
The first factor describes the distribution over the dilepton energy $q^0$, controlled by
the lifetime of the $B$ field $t_B$. 
 Since the energy dependence happens to be exponential, one may  call the inverse lifetime of the $B$ field an ``effective temperature".  Its value
 $T_{eff}=(2 t_B)^{-1}\approx 0.66 \,{\rm {\rm GeV}}$ (at RHIC) is rather high,  exceeding the actual initial QGP temperature $T_i\approx 0.35\, {\rm {\rm GeV}}$ which governs the thermal lowest-order dilepton production $\sim exp(-q^0/T_i)$  due to $\bar q q\rightarrow l^+l^-$ process, the dominant source of the IM dileptons.
 The second factor defining the dilepton momentum distribution comes from the stress tensor perturbations. It contains
 the temperature and the speed of sound $c_s\approx 1/\sqrt{3}$.  
   On top of these exponential
dependences, there are also important powers of $q^0$ ond/or dilepton mass, as well as various numerical factors,
which will be evaluated in what follows.

  \begin{figure}[h]
  \begin{center}
  \includegraphics[width=3cm]{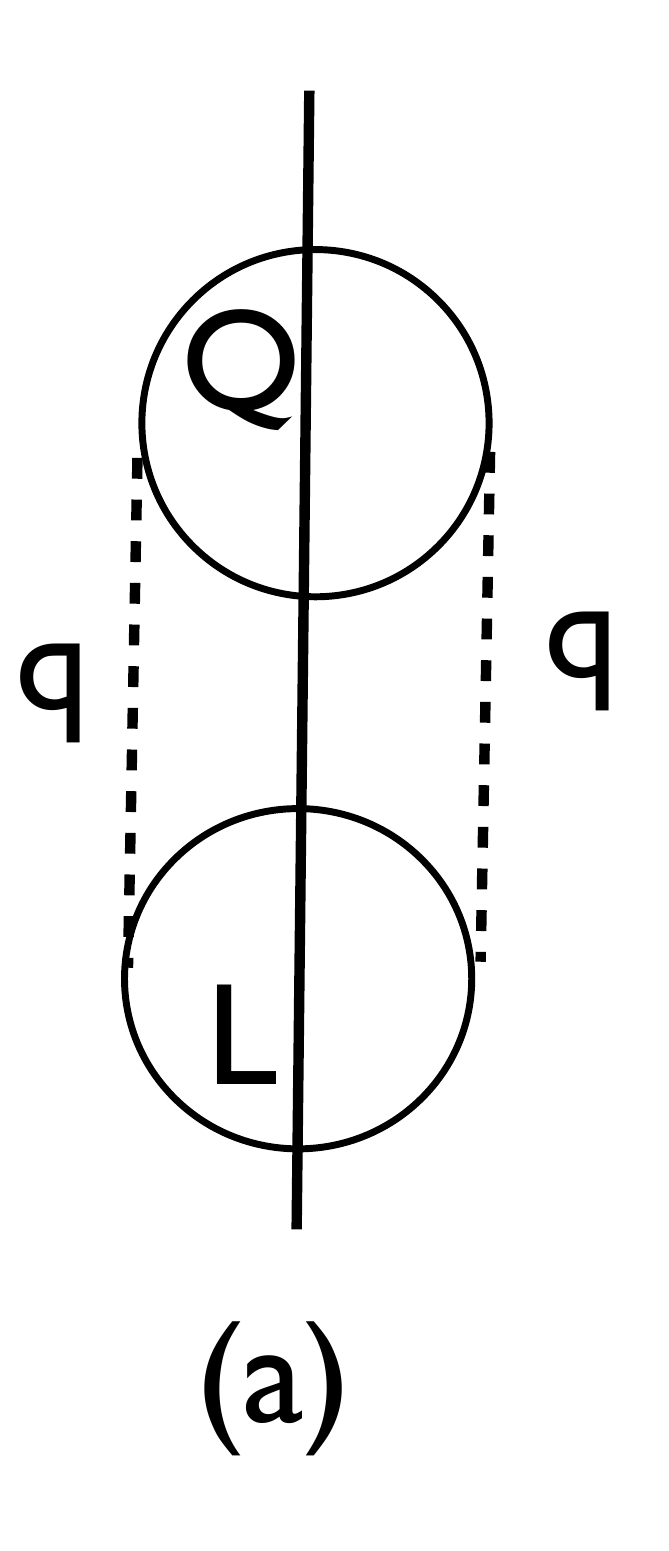}  \includegraphics[width=3cm]{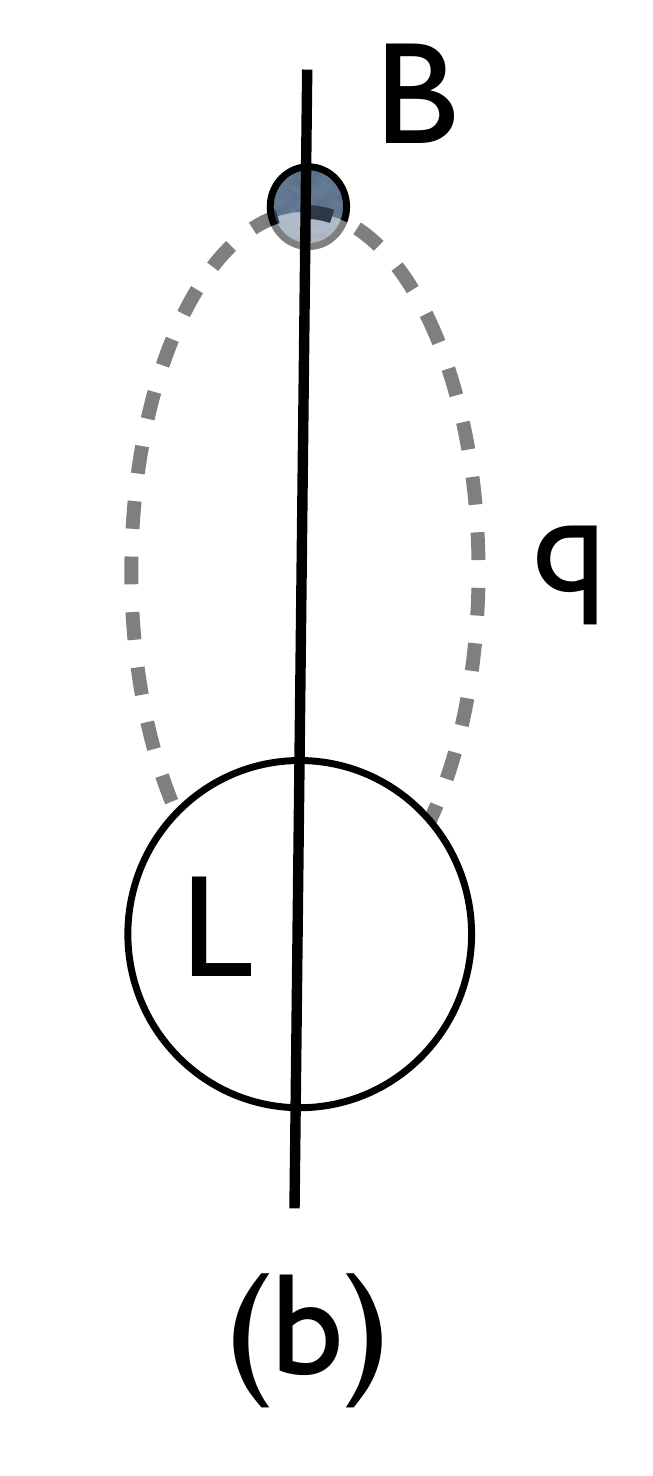}
  \caption{ (a) The diagram for the rate of the standard $\bar q q\rightarrow l^+l^-$ process. (b)
  Direct production of the dilepton from the magnetic field. Other notations are explained in the text.
  }
  \label{fig_diag}
  \end{center}
\end{figure}

  \begin{figure}[h]
  \begin{center}
  \includegraphics[width=7cm]{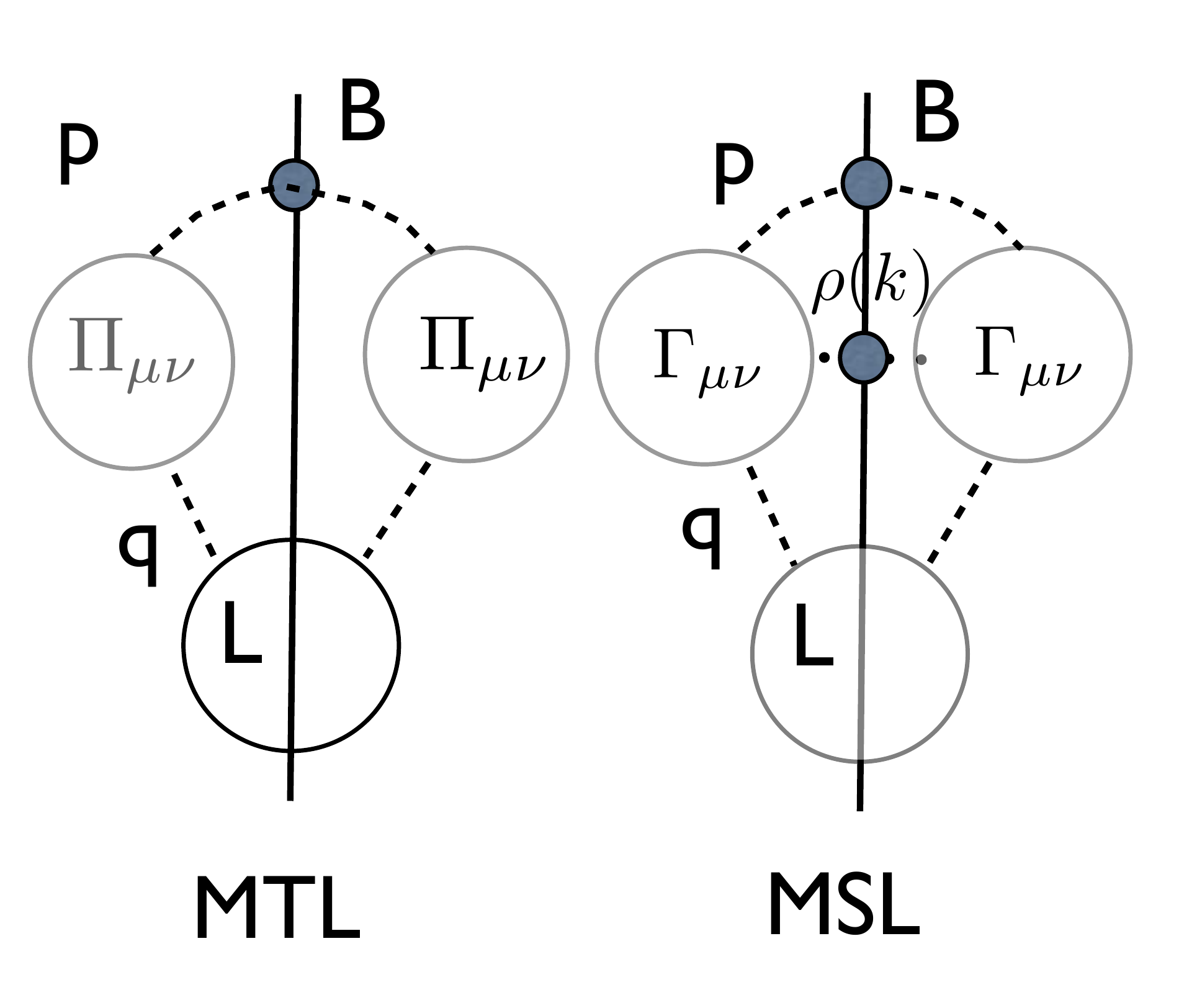} 
   \caption{ The schematic diagrams for the MSL and MTL processes. The lines marked by momenta $q,p,k$  are for virtual photon, magnetic field and phonons, respectively. 
 Other notations are explained in the text.
  }
  \label{fig_diag2}
  \end{center}
\end{figure}

\subsection{The outline of the calculation}
For completeness, let us start with the (well known since \cite{Shuryak:1978ij}) perturbative 
thermal rate in the zeroth order in strong interaction, due to $\bar q q\rightarrow l^+l^-$ process. 
In Fig.\ref{fig_diag}(a) this process is shown in the form of a ``unitarity diagram". It consists of the lepton loop $L$
and the quark loop $Q$, cut by the vertical line indicating the unitarity cut. All lines which are cut represent physical (on-shell)
quarks and leptons, with momenta denoted by $q_+,q_-$ and $l_+,l_-$, respectively. Two  dashed lines are the propagators of the
virtual photons, with momentum $q^\mu=q_+^\mu+q_-^\mu=l_+^\mu+l_-^\mu$. 

\vskip0.3cm
The MTL and 
 MSL rates are schematically given by the diagram shown in  Fig.\ref{fig_diag2}. Together with the lepton loop $L_{\mu\nu}$ unitarity cut,  it has two (uncut) copies of
the polarization operator $\Pi_{\mu\nu}$
 (or vertex $\Gamma_{\mu\nu}$) connecting  the virtual photon with momentum $q$
 with the photon with momentum $p$ and representing the external magnetic field.
  The MTL process has no sound but just a near-homogeneous medium with a stress tensor directly appearing
in the polarization operator $\Pi_{\mu\nu}$. The MSL case also has a
dotted line with momentum $k$; its unitarity cut includes the imaginary part of the  correlation function
 of two stress tensors in the medium.  

\vskip0.3cm

In the case of MTL process the transverse momentum of the dilepton pair is determined by the inverse size of the fireball and is thus very small.   
The transition from the MTL to MSL rate is done by replacing  the squared matter energy density in $\Pi$ by the integrated stress tensor spectral density,
  \be \epsilon^2 \rightarrow \rho(k) d^4 k; \ee 
  this will be explained in detail in Sec. \ref{sec_MSL}. In the MSL process, the dileptons get a substantial ``kick"  $q^\perp$ from phonons,  which transfer them into a more interesting kinematical region
 accessible to current experiments.

\section{The hadronic approach}
\label{sec_MTL_hadr}

\subsection{The effective Lagrangian}
We now move on to the calculation of the effective Lagrangian that couples two gluons to two photons via a quark loop. 
One phenomenological source of information about
    the virtual quark loop coupling to two photons comes from hadronic $\gamma \gamma$ decays; 
   for the scalar meson channel it has been used e.g. in \cite{Basar:2012bp}.
       Since in this paper we will deal with the stress tensor, we can use in a similar way
the      tensor meson decay to $\gamma \gamma$
to derive  an effective  coupling. This approach is similar to the well known vector meson dominance, and 
  is thus known  as the ``tensor meson dominance" \cite{Suzuki:1993zs}.  
  \vskip0.3cm

Before we do so, we would like to say a word of caution. The sigma meson, a chiral partner of the pion, represents the channel in which the strong attraction between quark and antiquark
    is known to exist. 
 Originally invented by Nambu-Jona-Lasinio (NJL) to describe the chiral symmetry breaking, those forces act at a scale $\sim 1\, {\rm {\rm GeV}}$.
They were later attributed to the 
instanton-induced effects and explain the pattern of chiral symmetry breaking and the $U_A(1)$ anomaly by providing repulsion in the $\eta'$ channel, for review see \cite{Schafer:1996wv}. These effects make $\sigma$ lighter but also more compact. 
There are evidences that the lowest multiplet of the vector mesons are also, in a way, waves in the chiral condensate: in particular the ``vector meson dominance"
in hadronic physics and dilepton production is well established.
There are no such evidences for the tensor $f_2$ channel. Thus the accuracy of the ``tensor meson dominance" remains unknown, and the analogy on which this subsection is based may be misleading. 
   \vskip0.3cm

  We define the effective Lagrangian for couplings between gluonic and photonic stress tensors and their traces as:
 \ba
 \mathcal L=g_T T_{\gamma\gamma}^{\mu\nu}  T_{\mu\nu}^{glue} +g_S F^2 T^\mu_\mu \label{Leff} ;
 \ea
 here $T_{\gamma\gamma}^{\mu\nu}= F^{\mu\alpha}F_{\alpha}^\nu-1/4g^{\mu\nu}F^2$ is the electromagnetic stress tensor and $T^{glue}_{\mu\nu}$ is the color traced gluonic stress tensor. 
 We fix the coupling $g$ in two steps. First, we use the tensor meson dominance; namely in the matrix element between
 the vacuum and photons, we keep only  the lightest tensor meson, $f_2(1270)$, term:
 \ba
 \langle0|\bar T^{\mu\nu}|f_2\rangle=m^2_f\,f_f\, \langle0|f^{\mu\nu}|f_2\rangle=m^2_f\,f_f\,\epsilon^{\mu\nu}.  \label{tensor_melement}
 \ea  
 Here $f^{\mu\nu}$ is the corresponding meson field and $\epsilon ^{\mu\nu}$is the polarization tensor that satisfies $\epsilon^{\mu\nu}=\epsilon^{\nu\mu}$, $\epsilon^\mu_\mu=0$, and $q_\mu\epsilon^{\mu\nu}=0$ for the $f_2$ momentum $q^\mu$. Notice that the polarization tensor has 5 degrees of freedom as expected for a spin-2 particle. The decay constant $f_f$ can be deduced using the input from the dominant  decay channel, $f_2\rightarrow \pi\pi$. Indeed, the effective Lagrangian for $f_2\pi\pi$ is
 \ba
 \mathcal L_{f\pi\pi}=g_{f\pi\pi} \partial_\mu \pi \cdot \partial _\nu \pi\, f^{\mu\nu}
 \ea
 and the tensor meson dominance relates the decay constant to the effective coupling as $g_{f\pi\pi}\,f_f=1$ in our normalization. The decay width is \cite{Suzuki:1993zs}
 \ba
\Gamma(f_2\rightarrow\pi \pi)=g_{f\pi\pi}^2{m_f^3\over 320 \pi}\left(1-4{m_\pi^2\over m_f^2}\right)^{5/2}.
 \ea
The most recent  values for the decay width and the mass of $f_2$ are  $\Gamma(f_2\rightarrow\pi \pi)\approx 157\ {\rm {\rm MeV}}$ and  $m_f\approx 1275\ {\rm {\rm MeV}}$ \cite{Beringer:1900zz}. Therefore we obtain the value of the decay constant $f_f\approx 108\ {\rm {\rm MeV}}$. 
 \vskip0.3cm

The next step is to fix the coupling of $f_2$ to electromagnetism. This can be achieved by analyzing the $f_2\rightarrow\gamma\gamma$ decay. From the $f\gamma\gamma$ effective Lagrangian
\ba
\mathcal L_{f\gamma\gamma}= g_{f\gamma\gamma} f_{\mu\nu} T_{\gamma\gamma}^{\mu\nu}, \label{Leff2}
\ea
we obtain the decay width as
\ba
\Gamma(f_2\rightarrow\gamma\gamma)=g_{f\gamma\gamma}^2{m_f^3\over 80\pi} .
\ea
Using the most recent value $\Gamma(f_2\rightarrow\gamma\gamma)\approx 3\ {\rm keV}$ from \cite{Beringer:1900zz}, we fix $g_{f\gamma\gamma}=0.014\ {\rm {\rm GeV}}^{-1}$. By connecting the matrix elements of Lagrangians (\ref{Leff}) and  (\ref{Leff2}), we obtain the effective $gg\gamma\gamma$ coupling
\ba
g_T={g_{f\gamma\gamma}\over m^2_f f_f}\approx(1.87\ {\rm {\rm GeV}})^{-4}.
\ea
In order to understand whether this coupling is large or small on the QCD scale, one can also redefine it by dividing by the electromagnetic coupling $\alpha_{em}$ and the relevant scale, taken to be the 
resonance mass
\be    \lambda_{h\gamma\gamma}=     g_{h\gamma\gamma} m_{h}/\alpha_{em}. \ee
The results are collected in the Table \ref{tab}, in which we also included those for the scalar channel from \cite{Basar:2012bp}.
Such dimensionless couplings are indeed numbers O(1).

\begin{table}[h]
\begin{tabular}{|c|c|c|c|c|c|  } \hline
              &  $m_h $           & $f_h$               &    $g_{h \gamma\gamma}$ & $\lambda_{h \gamma\gamma}$  & $g_i$ \\
& ${\rm GeV}$ & ${\rm GeV}$  & ${\rm GeV}^{-1}$ && ${\rm GeV}^{-4} $  \\    \hline
$\sigma\,  0^{++}$ & 0.55   & 0.10     & 0.02                & 1.5                                                     & 0.66    \\
f $2^{++}$               &  1.27 & 0.108   &  0.014               & 2.43                                                    &  0.08       \\ \hline
\end{tabular}
\caption{ (Color online)Comparison of the hadronic couplings to the $\gamma\gamma$ in scalar and tensor channels} \label{tab}
\end{table}
\vskip0.3cm
In what follows, we will however need an effective Lagrangian in the form of (\ref{Leff}). 
The relation we need, in order to get the scalar coupling analogous to (\ref{tensor_melement}), is the matrix element of the divergence of the dilatational current:  
 \be <0|\partial_\mu S_\mu |\sigma> = <0| T^\mu_\mu |\sigma>=m_\sigma^2 f_\sigma,  \ee
 that leads to $g_S=g_{\sigma\gamma\gamma} / m^2_\sigma f_\sigma $, similar to the expression for $g_T$. 
 Note however, that the relation between the gluonic stress tensor and the gauge field for tensor and scalar components is
 quite different; while the tensor is constructed out of the field strength, in a classical Maxwellian fashion, the scalar is classically zero and only appears 
 at the one loop order, through the scale anomaly relation
 \be T^\mu_\mu =  {\beta(g) \over 2 g} G^2\approx - g^2 {b\over 32 \pi^2} G^2    \label{eqn_anom} \ee
 with the first order beta function coefficient for QCD, \be b=(11/3)N_c-(2/3)N_f \approx 9.\ee  This form
will be important when we compare to the OPE expressions below.
 
\vskip0.3cm

 While the coupling in the scalar channel is
 several times larger than in the tensor one, the trace (scalar part) of the stress tensor in quark-gluon plasma is smaller than its tensor part.   Deviations from conformality in QGP are of the order
 \be{ T^{\mu\mu} \over T^{\mu\nu}}\sim  (1- 3c_s^2(T))^2 \sim \left({T_c\over T}\right)^2   \label{eqn_tmumu}
 \ee where 
 in the first equality we followed the arguments reproduced in Ref \cite{Basar:2012bp},
 $c_s^2=dp/d\epsilon$ is the sound speed squared, and the last equality comes empirically, from 
 the available lattice equation of state for $T>T_c$, see \cite{Pisarski:2006hz,Megias:2009mp}. For initial temperatures $T_i$ at RHIC/LHC the trace 
 is suppressed by about one order of magnitude.
Taking it into account,
together with couplings in the Table, one may conclude that the scalar and tensor parts of the glue will generate comparable electromagnetic effects.

\subsection{The MTL dilepton yield using the hadronic effective Lagrangian} 
 The Lagrangian (\ref{Leff}) directly couples the gluonic and electromagnetic stress tensors, and for the MTL case 
 the former is treated as the smooth mean describing the matter inside the fireball.
 At an early time and in the comoving frame, the stress tensor is dominated by gluons and has the form $T_g^{\mu\nu}=diag(\epsilon,p,p,p)$.
We also assume that the (collective) magnetic field has only one component, $B_3$.
This reduces the tensor part of the Lagrangian to  
\be L_{eff}= {g_T\over 2}  (\epsilon+p) B^A_3 B^B_3.
\ee
Going to the Fourier transforms of the fields one gets the MTL yield
\be 
dN^{MTL}= \left({g_T\over 2}\right)^2  (\epsilon+p)^2 \,\tilde{B}^2_3(q)\, q_2^2 \,{L_{11} \over M^4} \, PS(l^+,l^-),
\ee
where $q=l^++l^-$ and the lepton phase space
\be PS(l^+,l^-) = {d^3 l_+ \over (2\pi)^3 \epsilon_{+}}{d^3 l_- \over (2\pi)^3 \epsilon_{-}}  \ee
includes the spin factor 2$\times$2.
  \begin{figure}[t]
  \begin{center}
  \includegraphics[width=8cm]{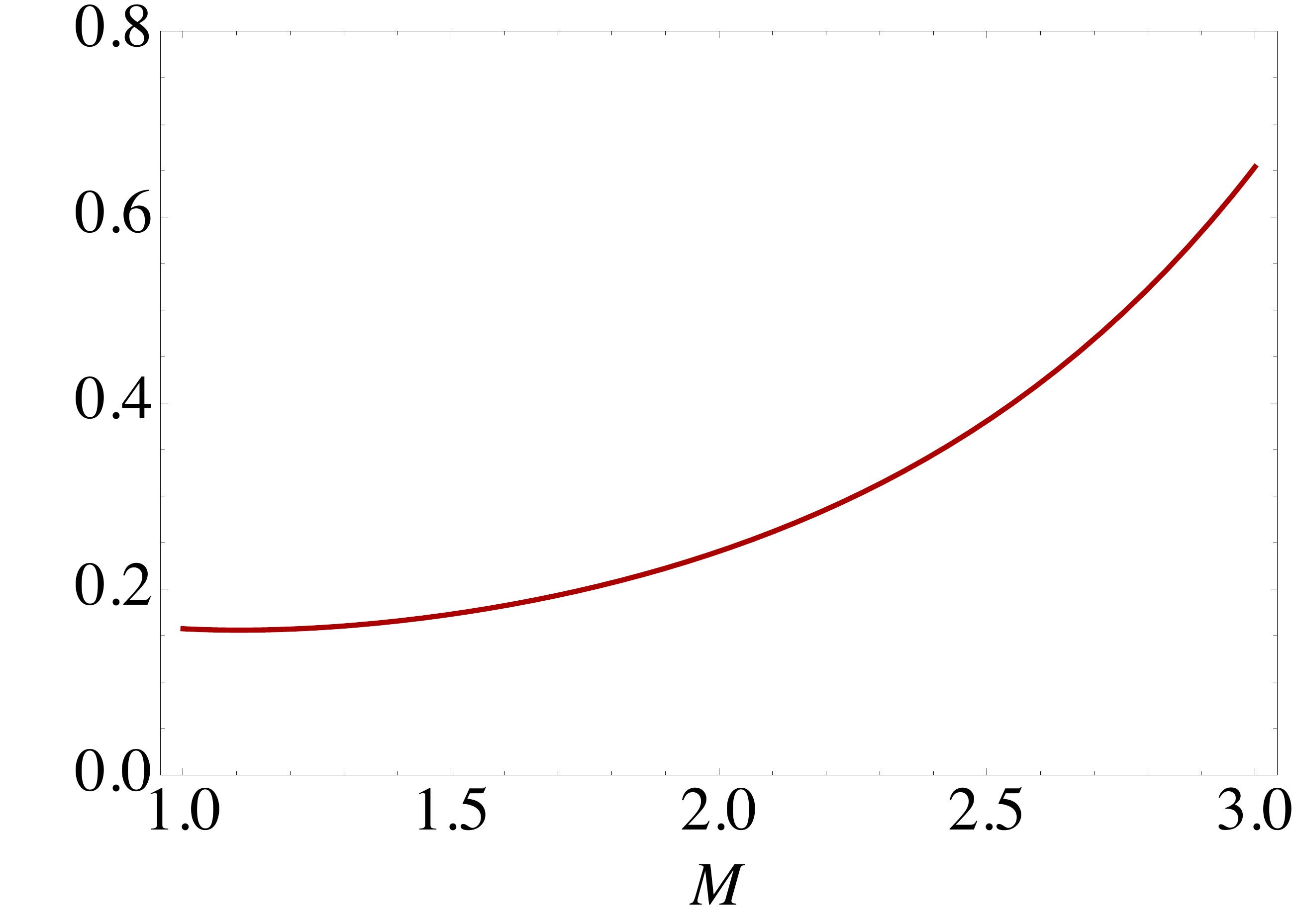}
  \caption{The ratio of the MTL yield of dileptons to that of lowest order annihilation process
  as a function of the dilepton mass $M$ (${\rm GeV})$ and small transverse momentum $q_2=1/R_2$. The coupling of photons to gluons is determined by the hadronic approach explained in Sec \ref{sec_MTL_hadr}. For the  values of the parameters see eqns (\ref{MTL_hadr_comp1}-\ref{MTL_hadr_comp2}).}
  \label{fig_MTL_qq}
  \end{center}
\end{figure}
\vskip0.3cm
Let us now compare the MTL process to the standard $\bar{q}q$ annihilation in QGP. Since the dilepton production in QGP and the processes we consider in this paper take place at different times, instead of comparing the rates, one should compare the $yields$. The temperatures are also quite different, due to two reasons. The first is that
the system is expanding and thus $T$ is time dependent. We take care of the time dependence by assuming a one-dimensional Bjorken expansion and entropy conservation.
The other reason is the different number of degrees of freedom (NDF) in ``hot glue" and QGP:
\ba 
NDF_g & = & 2\times 8 =16,\\
\label{MTL_hadr_comp1}
NDF_{qgp}& = & 16+N_c N_f  \times2\times(7/8)\approx 47. 
\ea 

 The relation between the two temperatures is then 
 \be T_g = T_{qgp}(NDF_g\times\tau_g/(NDF_{qgp}\tau_{qgp})^{-1/3} ;
 \ee
 the typical values we will use for comparison are 
\ba \tau_g=t_B=0.15\ fm,&\quad&   T_g = 1\, {\rm {\rm GeV}} \\
 \tau_{qgp}=2 \, fm, &\quad& T_{qgp}=0.3\,  {\rm {\rm GeV}} . 
 \label{MTL_hadr_comp2}
 \ea
 Note that the quark suppression fugacity  in the comparison so far is put  to $\xi=1$. However, if the
 suppression is real and the ``hot glue" scenario is correct, the ratio $T_{qgp}/T_g$ is to increase by a factor $1/\xi^2$
 and may actually be around 1.

For particular kinematics, in which the dilepton transverse momentum is very small ($p_2=1/R_2$)
and for parameters listed above, we plot the ratio of the yields in Fig.\ref{fig_MTL_qq}. 

\section{The OPE approach: Deep-inelastic scattering  on a  gluonic medium } \label{sec_Pi}
In this section, we calculate the effective $gg\gamma\gamma$ Lagrangian using the tools of the Operator Product Expansion (OPE), analogous to the study of Deep Inelastic Scattering (DIS).
DIS is a traditional tool used since the early days of QCD to describe 
the interaction of virtual photons with quarks located inside the nucleons and nuclei. 
Its main characteristic is the assumption that  the virtual photons involved have a virtuality scale $q^2$ which is
large compared to the virtuality scale inside the target.
\vskip0.3cm

By the gluonic medium, we mean the medium produced in early stages of the heavy ion collision. 
``Hot glue" is a term \cite{Shuryak:1992bt} for thermally equilibrated glue, without chemical equilibration to full QGP with quark-antiquark pairs.
``Glasma" is the term \cite{Lappi:2006fp} that describes an ensemble of gauge fields which is out of equilibrium and,
 following the argument of McLerran and Venugopalan \cite{McLerran:1993ni},
  has large occupation numbers and   can be described by classical, coherent Yang-Mills fields that emerge from random sources.
\vskip0.3cm
There are numerous examples of effective Lagrangians induced by fermion loops that range from the QED Heisenberg-Euler Lagrangian to
the lagrangians induced by heavy quark loops in the Standard Model. For multi-gluon processes considered in \cite{Chiu:2012ij}, these lagrangians however  have not yet been calculated, and only generic estimates for them have been used so far. 
\vskip0.3cm
Here we derive an effective Lagrangian of $\gamma\gamma$ interaction with certain gluonic operators of the $gg$ type. We start with  the simplest kinematics,
in which the photon momentum   $q$ is very large compared to that of the glue. In this case,
the effective Lagrangian reduces to the calculation of the polarization tensor \be \Pi_{\mu\nu}=<J_\mu(x) J_\nu(0)>|_G \ee
evaluated on a background with a ``soft" colored field $G_{\mu\nu}$. The distance $x\sim 1/q$ is considered small,
and therefore the calculation represents an example of the OPE, expressing a bi-local (non-local in general)
expression as a series in powers of $x$ with coefficients representing local operators at point $0$. 
\vskip0.3cm
On general grounds (gauge invariance and current conservation), the polarization tensor is transverse to the vector $q_\mu$. This fact, as well as the fact that matter has a rest frame,
defined by the unit 4-vector $n_\mu$, allows for the traditional decomposition into two structure functions (analogous to DIS on a nucleon)
\be  \Pi_{\mu\nu} = P^1_{\mu\nu}(q) W_1+P^2_{\mu\nu}(q)  W_2  \label{eqn_P1P2}  \ee
$$ P^1_{\mu\nu}(q)=-(g_{\mu\nu}-q_\mu q_\nu/q^2) $$
$$ P^2_{\mu\nu}(q) =\left(n_\mu-{(n.q) \over q^2}q_\mu  \right)\left(n_\nu-{(n.q)\over q^2} q_\nu \right)
$$

However unlike in DIS, the kinematics here correspond to a time-like $q^2$ with a positive dilepton mass. In general, there are two rest frames: that of the medium $n_\mu=(1,0,0,0)$
and that of the dilepton. In general they do not coincide and this fact produces the second structure function.
 
\vskip0.3cm
One similar example in QCD is the celebrated derivation of the scalar gluon operator $G^2$ (known as the gluon condensate) correction to the polarization tensor,
 by Shifman,Vainshtein and Zakharov \cite{Shifman:1978bx}, which involved a rather complicated diagrammatic calculation. (Of course in vacuum there is no matter and no $n_\mu$
 vector, so there is only one transverse structure made of the $q$ vector and only one structure function $W_1$.)
We will however follow a different path, developed by Vainshtein and Shuryak  \cite{Shuryak:1981pi}, in connection to power corrections to DIS on the (polarized) nucleon.
For a pedagogical introduction see e.g. \cite{Shuryak:1988ck}.  
\vskip0.3cm
One element of this calculation is the use of the so called fixed-point gauge
\be x_\mu A_\mu(x)=0,\ee
 invented by Fock, Schwinger and perhaps others. In this gauge $A_\mu(0)=0$ and next order terms in $x$ expansion can be written as covariant derivatives
 of the field strength
 \be A_\mu(x)=\sum_{k=0} {x_\nu x_{\alpha_1}\ldots  x_{\alpha_k}  \over k !  (k+2)} D_{\alpha_1} \ldots  D_{\alpha_k} G_{\mu\nu}(0) 
 \label{eqn_Amu}
 \ee 
  In coordinate representation, the polarization operator is simply a fermionic loop
 \be  \Pi_{\mu\nu}(x) = Tr (\gamma_\mu S_G(0,x) \gamma_\nu S_G^+(x,0))\ee
 where the main ingredients -- the propagators -- are  to be calculated in a background field $G$, as indicated by the subscript. 
 The OPE expansion proceeds in powers of distance $x$, that is assumed small compared to the typical variation scale of the field and its derivatives. 
 In what follows, we will assume that the gauge field $G$ is approximately constant and therefore neglect its gradients. 
 \vskip0.3cm
 
 Let us present the calculation for one flavor of massless quark with a unit charge. 
  The propagator in the fixed-point gauge has been calculated in  \cite{Shuryak:1981pi} in momentum space to the order that we need.   We only include the terms that involve one and two field strengths, and ignore the covariant derivatives of the fields (because we assume it to be constant for now): 
 \ba &&\hskip-0.5cmS_G(q)={1\over \slashed{q}} - {g \over 2 q^4} q_\alpha \tilde{G}_{\alpha\beta} \gamma_\beta \gamma_5 \nonumber\\
&&\hskip-0.5cm-{g^2 \over 2 q^8}\slashed{q} q_\alpha G_{\alpha\beta} G_{\beta\gamma} q_\gamma+{g^2 \over 4 q^8} q^2 q_\alpha \{ G_{\alpha\beta} , G_{\beta\gamma} \}_+ \gamma_\gamma \nonumber\\
 &&\hskip-0.5cm- {g^2 \over 4 q^8}q^2 q_\alpha [ G_{\alpha\beta} , G_{\beta\gamma} ]_- \gamma_\gamma +O({1 \over q^6}), \label{eqn_prop}
 \ea
  where
  $\slashed{q}=q_\mu \gamma_\mu$ and the gluon field strength is assumed to be a
   color matrix $G_{\mu\nu}= G_{\mu\nu}^a t^a$, where $t^a$ are the SU(3) Gell-Mann matrices.  
 \vskip0.3cm
 
 The symmetric part of the term quadratic in the field strength can be expressed as  $\frac{g^2}{2 q^8}T_{\alpha\beta}(\slashed{q}q^\alpha q^\beta-q^2 q^\alpha\gamma^\beta)$, where $T_{\alpha\beta}$ is the traceless gluonic energy momentum tensor:
 \be
T_{\alpha\beta}=\frac{1}{2}\{G_{\alpha\mu}, G_\beta^{\,\,\,\mu}\}-\frac{1}{4}g_{\alpha\beta}G_{\mu\nu}G^{\mu\nu}
\ee
We remind the reader that this is still a matrix in the color space. The fermion loop leads to a color trace, and projects on the colorless 
 part, which is denoted by a bar above the operator
 \be  \bar{T}_{\alpha\beta}\equiv{1 \over 2}Tr_c T_{\alpha\beta} \label{glue_t} \ee
 
 In momentum space, the loop is an integral over the convolution of the two propagators with different arguments, which is complicated.
However going to coordinate representation, one finds simpler Feynman rules, in which the loop is just a product of two propagators $S(x)$ $without$ the integral.
Following this idea, we rewrite the propagator in the coordinate space:

\ba
 S_G(x)&=&\frac{i \gamma_\mu x^\mu}{2\pi^2 x^4}-i\frac{g}{16\pi^2 x^2}x^\mu \gamma^\nu \gamma^5\tilde G_{\mu\nu}\nonumber\\
 & -&i\frac{ g^2}{384\pi^2x^2}\gamma_\lambda x^\lambda x^\mu x^\nu T_{\mu\nu}\nonumber\\
&+&i\frac{g^2}{96 \pi^2}x^\mu \gamma^\nu T_{\mu \nu} \ln x. \label{Spos}  \ea

In the expression of the position space propagator above, we did not include the commutator term because being a traceless matrix, it will not contribute to the self energy up to second order in field strength.  In position space, the self energy $\Pi_{\mu\nu}$ is simply a product of two propagators:
\be
\Pi_{\mu\nu}(x)=Tr[\gamma_\mu S(x) \gamma_\nu S^\dagger(-x)],
\ee
where the trace is both over color and Dirac structure. Using the result (\ref{Spos}) for the position space propagator, we obtain
\ba
&&\hskip-0.5cm\Pi_{\mu\nu}(x)=\frac{-3}{\pi^4 x^8}(2x_\mu x_\nu-g_{\mu\nu} x^2) \nonumber\\
&-&\frac{g^2}{64\pi^4}\frac{x^\alpha x^\beta}{x^4} tr_c[\tilde G_{\mu\alpha}\tilde G_{\nu\beta}+\tilde G_{\nu\alpha}\tilde G_{\mu\beta}]\nonumber\\
&+&\frac{g^2}{64\pi^4}\frac{x^\alpha x^\beta}{x^4} tr_c[g_{\mu\nu}\tilde G_{\alpha}^{\,\,\,\lambda}\tilde G_{\beta\lambda}]\nonumber\\
&-&\frac{g^2}{48\pi^4}\frac{\ln (m^2x^2)}{x^4}tr_c[x_\mu x^\alpha T_{\alpha\nu}+x_\nu x^\alpha T_{\alpha\mu}] \nonumber\\
&+&\frac{g^2}{48\pi^4}\frac{\ln (m^2 x^2)}{x^4}tr_c[g_{\mu\nu}x^\alpha x^\beta T_{\alpha\beta}] \nonumber\\
&+&\frac{g^2}{96\pi^4x ^6}(2x_\mu x_\nu x^\alpha x^\beta -g_{\mu\nu}x^2x^\alpha x^\beta) tr_c[T_{\alpha\beta}]. 
\label{selfx}
\ea
The color trace  $tr_c(t^m t^n)=2\delta^{mn}$ will lead to an extra factor of 2 in the quadratic terms.  Evaluating the color trace and going back to momentum space, we get
  \ba
 &&\hskip-0.5cm\Pi_{\mu\nu}(q)=-\frac{1}{4\pi^2}(q_\mu q_\nu-g_{\mu\nu}q^2) \ln (q^2/4m^2)\nonumber\\
 &+&\frac{g^2}{4\pi^2}\frac{q^\alpha q^\beta}{q^4} \tilde G^a_{\mu\alpha}\tilde G^a_{\nu\beta}\nonumber\\
 &+&\frac{g^2}{36 \pi^2q^4}\left(q^2 \bar{T}_{\mu\nu}-(q_\mu \bar{T}_{\nu\alpha}+q_\nu \bar{T}_{\mu \alpha})q^\alpha\right)\nonumber\\
 &+&\frac{g^2}{36 \pi^2q^4}(6q_\mu q_\nu-5g_{\mu\nu}q^2)\frac{q^\alpha q^\beta}{q^2}\bar{T}_{\alpha\beta}\nonumber\\
 &+&\frac{g^2}{6\pi^2}\frac{\ln (q^2/\mu^2)}{q^4}\left(q^2\bar{T}_{\mu\nu}-(q_\mu \bar{T}_{\nu\alpha}+q_\nu \bar{T}_{\mu \alpha})q^\alpha\right)\nonumber\\
 &+&\frac{g^2}{6\pi^2}\frac{\ln (q^2/\mu^2)}{q^4}\left(g_{\mu\nu}q^\alpha q^\beta \bar{T}_{\alpha\beta}\right).\
  \ea
 Note that the CP-odd structure $A\sim  \vec{E}.\vec{B}$ does not appear in our effective interaction, to the order we consider. 
As for the scalar part, one can now recall the scale anomaly (\ref{eqn_anom}) to $g^2$ order and rewrite it in terms of trace $T^\mu_\mu$.  

Using the direction averaged parameterization for the gluon field, we arrive at the desired form of the self energy:
\begin{widetext}
\ba  \label{eqn_pimunu}
&&\hskip-0.5cm<\Pi_{\mu\nu}>=-\left( q^2 \frac{\ln (q^2/\mu^2)}{4\pi^2} 
 + {2\over 3b q^2}  T^\alpha_\alpha \right) P^1_{\mu\nu}
+\frac{ g^2 C}{9 \pi^2 q^2}\left(22+24  \ln ({q^2 \over \mu^2})\right)P^2_{\mu\nu}\nonumber\\
&-&\frac{g^2C}{9\pi^2 q^2} \left(5-2 \frac{(n.q)^2}{q^2} +12\left[1+2 \frac{(n.q)^2}{q^2} \right] \ln ({q^2 \over \mu^2})\right)P^1_{\mu\nu},\nonumber\\
\ea
\end{widetext}
where the two projectors $P^1_{\mu\nu}$ and $P^2_{\mu\nu}$ are defined in (\ref{eqn_P1P2}).
 The first term  is the usual QED beta function, the second is the famous SVZ result,
 used for evaluation of the effect of vacuum gluon condensate on the correlator. Of course,  in our case the operator should be evaluated over the medium, not the vacuum. 

 Observation shows that while the coefficients of the invariant $C$ (proportional to the stress tensor, see Appendix \ref{app_glue} for the definition) are proportional to $g^2/4\pi^2=\alpha_s/\pi $ =O(1/10) and are parametrically suppressed, the trace term is only numerically suppressed. However,
  the suppression of the trace term by the beta function parameter $1/b=1/9$ is of similar smallness. Thus at the level of coefficients,
  the two channels seem comparable. Yet the near-conformality of the sQGP  suppresses the magnitude of the trace further, 
  as indicated in (\ref{eqn_tmumu}) by an additional order of magnitude.
 We expect the scalar combination be in fact subleading,  and thus will keep the stress tensor only in 
 estimates to follow.
 
\section{The dilepton yields for MTL processes} \label{sec_MTL}
\subsection{The dilepton rates}
\label{subsec_rate}
In this section, we gather all the ingredients of our effective Lagangian to calculate the dilepton yield. In coordinate representation it is a nonlocal action
\be S_{eff}= \int d^4x  d^4y A^A_\mu(x) \Pi^{\mu\nu}(x-y) A^B_\nu(y)  \ee
 which is simplified in momentum representation.
\vskip0.3cm

We start
with the MTL case, in which there are no phonons and the stress tensor of the medium is just a smooth collective energy density.
The effective action for MTL is given by the convolution of the polarization tensor, derived above, with two photon fields. Therefore, 
in the effective diagram  shown in Fig.\ref{fig_diag}(b) there is no line marked by momentum $k$.
In this case there is only one loop 4-momentum $p^\mu=q^\mu$
on which the polarization tensors  $\Pi^{\mu\nu} $ depend, as already derived in (\ref{eqn_pimunu}).  The same momentum is the argument
of two  ``propagators" of the photons, including dilepton bracket and magnetic source. The dilepton rate of the MTL process is
\be \label{eqn_rate}
dN_{l+l-}   = {d^4 q d^4 x \over (2\pi)^4} \,   [<A^A_\mu A^A_{\mu '}> \Pi^{\mu'\nu'}  
<A^B_{\nu'} A^B_{\nu }> \Pi^{\mu\nu} ]. \ee
The labels A and B correspond to the
virtual photon producing dileptons and to the magnetic field, respectively. 
\vskip0.3cm

The angular brackets imply that these quantities are ``propagators" of the gauge fields.
In ordinary diagrams they are defined via Fourier transforms of the fields
\be \tilde A(k)=\int d^4x e^{ikx} A(x) \ee
and their bilinears are replaced with propagators as follows
\be \tilde A(k) \tilde A(k') \rightarrow <\tilde A_k \tilde A_k> (2\pi)^4\delta^{(4)}(k-k'). \ee
The mass dimensions of $A, \tilde A, <\tilde A_k \tilde A_k>$ are 1, -3, -2, respectively.
The dimension of $\Pi$ is 2, and sequence of propagators and $\Pi's$ in equal number is dimensionless, so is
the phase space in (\ref{eqn_rate}).
\vskip0.3cm
However, the propagators with the unitarity cut are slightly different, 
and include certain spectral densities of on-shell objects. The one for the dilepton is
\be
  <A^A_\mu A^A_{\mu '}>  =  
 4{L_{\mu\mu'} \over (2\pi)^6 M^4} {d^3 l_+ \over 2 \epsilon_{+}}{d^3 l_- \over 2 \epsilon_{-}} (2\pi)^4 \delta^4(l_++l_--q),
\ee
in which the (polarization summed) lepton tensor corresponds to the usual fermion loop; the factor 4 in front is from spin summation.
Since mass dimension of $L_{\mu\mu'} $ is 2, it has the same dimension as the propagator. Note that in the center of mass frame
of the dilepton, the 6-4=2 remaining integrals, after the delta function is fully used, are dimensionless and taken over the solid angle of the lepton directions in this frame.
One cannot however easily perform these integrals in general, since the indices of $L_{\mu\mu'} $ are coupled to others in the diagram.
\vskip0.3cm

The MTL dilepton yield is thus given by 
\be dN^{MTL}=|\tilde A^B(l_++l_-)|^2 {(\Pi L \Pi)_{11} \over M^4} PS(l^+,l^-). \ee
 The $\Pi L \Pi$ is a shorthand notation, in which the convolution in the indices of the polarization and lepton tensors are assumed, and the subscript $11$ indicates the values of the remaining outer indices; they are longitudinal  due to our selection of the $A^B$ field gauge. 
We have also obtained the general expression for the yield, but it is too lengthy to be given here.

\subsection{Dilepton kinematics}
 Let us now define the  kinematics of the dileptons and make some assumptions for simplicity, as the 6-dimensional distribution
is too complicated to study. 
 We first assume that the dilepton pair has no
longitudinal momentum in the medium, $q_1=0$, but may have a transverse one.  
The lepton 
 momenta can then be defined in terms of two vectors, the sum (still called $q$) and a new vector $m$:
 \ba  l_{\pm}[1]=\pm {m \over 2} \cos \theta, \nonumber \\ 
   l_{\pm}[2]={q_t \over 2} \cos\psi \pm { m  \over 2}\cos\phi \sin\theta, \nonumber \\ 
  l_{\pm}[3]={q_t \over 2} \sin\psi \pm { m \over 2} \sin \phi \sin\theta.
 \ea
  Note that we now have 3 angles, $\theta,\phi,\psi$. The angle $\theta$ is associated with the  polarization of the virtual photon, and $\psi$ is associated with 
  the total dilepton direction of motion in the transverse plane. The so called
  ``elliptic flow" is thus a moment associated with $\cos(2\psi)$. $\phi$ is the relative angle of the leptons in the transverse plane.
 \vskip0.3cm
 
 Since the mass of the leptons is negligible, one can consider the lepton energies to be simply the modulus of their momenta.
The general expressions for these energies are somewhat complicated. However if one 
 assumes that the dilepton transverse momentum is small compared to its mass,  i.e. 
 \be q_t \ll m, \ee 
 many expressions are simplified greatly, in particular the invariant mass 
 \be
M= \sqrt{-q^2}=m+ O( q_t^2 /k) \approx m
 \ee
 (thus the name of the vector). This kinematical window is where the main contribution to dilepton production  in the IM range. 
 \vskip0.3cm
 
For $q_t=0$ we are left only with the vector $m$. In this case
\be 
(\Pi L \Pi)_{11}^{{\bf q}=0} = g^4 C^2 { 8\sin^2\theta \over 9 \pi^4 M^2} \ln^2({M^2\over \tilde \mu^2}),
\label{zero_qt}
\ee
which depends only on one angle $\theta$ in a simple way, namely $\sin^2\theta$. This happens because in this case $n_\perp=0$
 and the second structure function in the DIS on matter disappears; it corresponds to a purely $longitudinally$ polarized
 virtual photon. We also defined $\tilde \mu\equiv e^{7\over24}\mu$.

 \subsection{Comparison of OPE estimates to the hadronic estimates for the MTL rates}
 
\begin{figure}[t]
  \begin{center}
  \includegraphics[width=8cm]{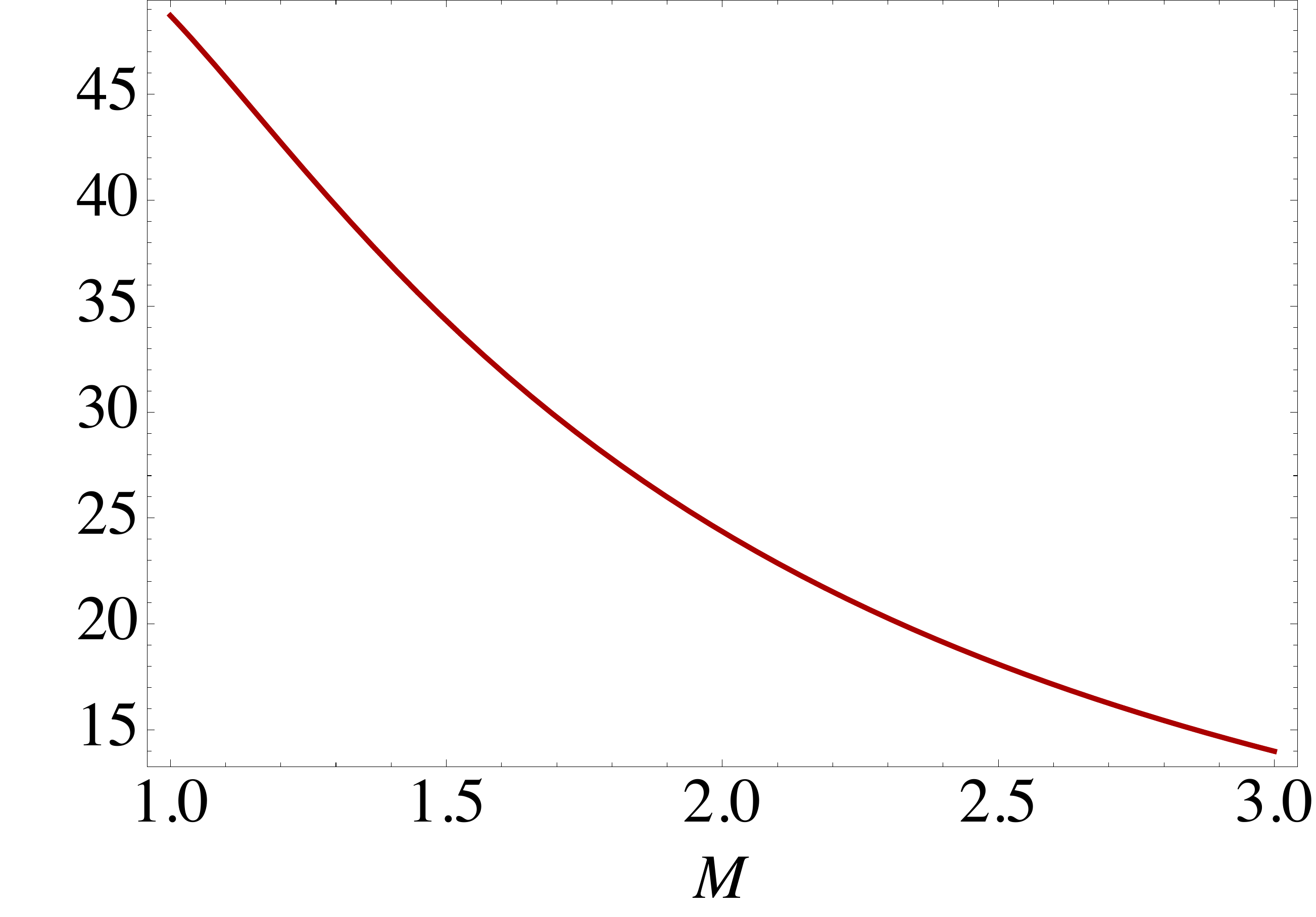}
  \caption{The ratio of the effective $gg\gamma\gamma$ coupling of the OPE approach to that of the ``tensor dominance", as a function of the dilepton mass $M$ ({\rm GeV}). For values of the parameters see text.}
  \label{fig_OPE_hadr}
  \end{center}
\end{figure}

 The resulting dilepton yield obtained from the (perturbative) OPE in the preceding section should be compared to
 the one obtained by tensor dominance in Sec. \ref{sec_MTL_hadr}. The matter density, leptonic phase space and magnetic field
 strength factor cancel out, and the ratio of the effective coupling of OPE to that of the tensor dominance takes a rather simple form
 \be 
 {\text{OPE coupling} \over \text {tensor  dominance coupling}}= {8 \alpha \alpha_s \over 3} { R_2^2 \ln(M/\tilde{\mu}) \over g_T M^2} 
 \ee
 which is plotted in Fig.\ref{fig_OPE_hadr} for the size of the fireball $R_2=6 fm, \alpha_s=1/3$, and $\tilde{\mu}=0.5\, {\rm GeV}$. 
 One can see that the two approaches yield significantly different predictions, especially at  
 the lower edge of the intermediate mass dilepton (IMD) range, $M\sim 1\, {\rm GeV}$. 
 It is known that the tensor meson dominance is not as accurate as for instance the vector meson one. We thus think  
 that at least at the high end of the IMP range,   $M\sim 3\, {\rm GeV}$,
 perhaps approaching  the domain of perturbative QCD, the OPE prediction should be reasonably reliable. 
 \vskip0.3cm
 
 Let us however make a  warning about the last statement:
 the boundary of perturbative QCD remains a hotly debated open issue. Not intending to entering it, we  remind that
 for example, such quantities as the pion and nucleon electromagnetic form factors, which were recently 
 measured experimentally at comparable scale of momentum transfers at JLAB (see e.g. \cite{Horn:2006tm}), are not in agreement with pQCD predictions.
 How much the situation improves at $T\sim 2T_c$ in the QGP phase that we discuss remains unclear.

\subsection{MTL process in experiment?}
Let us now discuss the MTL process in more detail.
 Theoretically, it is much simpler than MSL. For example, simpler kinematics transforms 
  the polarization tensor $\Pi_{\mu\nu}$ into the forward-scattering amplitude. 
  This makes, for example, checking  the gauge invariance much simpler. 
  \vskip0.3cm
  
  However the price one has to pay for such
simplicity is high: the MTL dileptons must have rather small transverse momenta $q_\perp\sim p_\perp \sim 1/R_{fireball}$, and this is the kinematical domain populated by numerous background sources of the dileptons. For example, a critical reader might have already wondered if the presence of hadronic matter is really necessary,   
since there exists a  simpler process shown in  Fig.\ref{fig_diag}(b) in which a
(time-dependent) collective magnetic field of the two passing ions can create the dileptons directly. 
Since the magnetic field extends well beyond the ions themselves, these processes should happen as well at large distances
from both ions, including  collisions in which the two ions pass at large impact parameters
without strong interactions. 
 There exist also higher order processes (not shown), in which two Weizseker-Williams photons collide
and produce dileptons, or bremstrahlung accompanying a Coulomb scattering, which have huge cross sections 
due to powers of the ion charge $Z$ appearing due to coherence of the field, see \cite{Baltz:2007kq} for review. On top of these textbook processes,
there are also ``semi-coherent" processes identified in \cite{Staig:2010by}. 
All of these processes can create dileptons which extend down to very small transverse momenta of the order of the lepton mass $q_\perp\sim m_l$.
(Note that we do not discuss  here the transverse momenta of the individual leptons, but of the pair.) 
  \vskip0.3cm
  
  Experimental studies of the dilepton production in this kinematics have $not$ been performed so far. For example, the standard setting of the PHENIX detector magnetic field
at RHIC at present simply cuts off these dileptons, observing only di-electrons with
 $q^\perp> 100\, {\rm MeV}$ or so. At the boundary of this kinematical region, an enhancement of the rate has been observed,
 but its nature has not yet been clarified. (It can be done e.g. by lowering the magnetic field.)
 \vskip0.3cm
 For these reasons, the evaluation of the MTL rate is at present of purely academic interest, and we will not  go further into a detailed study of the
 the diagram (b), or of the multiple background processes mentioned. Instead, we will concentrate on the MSL processes that, as we will soon see, do contribute substantially to the observable photon and dilepton production.

  \section{The sound-to-light conversion: MSL process}
   \label{sec_MSL}
  \subsection{The kinematics of MSL}

In the calculations of the MTL process above, we assumed that the smooth (gluonic) stress tensor 
is constant, and carries no momentum.  
Therefore  the dilepton 4-momentum coincided with that of the $B$ field, namely $q=p$.
\vskip0.3cm

 Now we would like to 
include also the effect of the perturbations of the stress tensor $\delta T_{\mu\nu}(k)$ -- called here 
phonons for brevity -- which carry a 4- momentum $k$. We remind that the momentum conservation now reads  $q=p+k$.
As a result,  what used to be the ``forward scattering" on the medium, described by the polarization tensor $\Pi_{\mu\nu}(q)$,
has to be lifted to a (non-forward) vertex  function $\Gamma_{\mu\nu}(p,q)$. Squares of the stress tensor will
be lifted to their correlation functions (see below).
\vskip0.3cm
QED gauge invariance
requires transversality of both photons
\be 
q_\mu \Gamma_{\mu\nu}(p,q) =0, \,\, \Gamma_{\mu\nu}(p,q) p_\nu=0.
\ee
 This can be enforced by a straightforward modification of two projectors defined in
(\ref{eqn_P1P2}) into the following form
\ba
\tilde{P}^1_{\mu\nu}&=&  P^1_{\mu\alpha}(q) P^1_{\alpha\nu}( p)  \nonumber \\
 \tilde{P}^2_{\mu\nu}&=&\left(n_\mu-{(n.q) \over q^2}q_\mu  \right)\left(n_\nu-{(n.p)\over p^2} p_\nu \right)
\ea
Unfortunately, there exists {\em one more} structure which satisfies both transversality conditions, namely
\be 
P^3_{\mu\nu}={ p_\mu q_\nu \over (p q)}+  { p_\nu q_\mu (qp) \over p^2 q^2} - { q_\nu q_\mu \over q^2} -{ p_\nu p_\mu \over p^2}
\ee
which vanishes at $p=q, k=0$. Thus the corresponding structure function cannot be recovered from the $k=0$ calculation.
\vskip0.3cm

Although we will not need to do so, for reasons to be explained shortly, it is instructive to note how
one should calculate such terms. In the selected gauge, the expression for $A_\mu$ (\ref{eqn_Amu})
contains the operators   of the structure $D_{\alpha1} \ldots  D_{\alpha k}  G_{\mu\nu}$. The derivative $D$ appears if the gauge field is not constant,
and can be identified with the momentum $k$ of the glue. At a given order $n$ of $k$, one needs to keep the same number of derivatives.
At the end, all structures proportional to the $n^{th}$ power of $k$ can be recovered.

\vskip0.3cm
However, the calculation  can be greatly simplified in the particular kinematics which we will use.
Note first, that if the small  3-momentum of the $B$ field (related to the inverse ion size) is ignored, $p_\mu$ has only one non-zero component $p_0\neq 0$. Furthermore, for any rapidity of the produced dilepton, the largest contribution comes from the part of matter comoving with it with the same rapidity. Focusing on the mid-rapidity $y=0$ and 
 the matter at rest  $n_\mu=(1,0,0,0)$,  we observe that the last bracket in $\tilde{P}^2_{\mu\nu}$
vanishes. 

\vskip0.3cm
Moreover, for the magnetic field, we use the gauge where the only non vanishing component of the potential for magnetic field, $A_\mu^B$,  is along the
longitudinal direction $\mu=1$. If the dilepton and phonon have both zero longitudinal momenta $q_1=0,p_1=0$,     
we find that  the projector $P_3$ does not contribute to the rate. We thus find that
the only contributing term is the projector $\tilde{P}^1_{\mu\nu}$, since it contains $\delta_{11}$. As a result, the longitudinal index is simply  passed to the
lepton tensor, and we only need $L_{11}$. This makes the ``parent" of the leptons -- the virtual photon -- to be $longitudinally$ polarized, as we already noticed before.
The resulting expansion for the $L_{11}$ up to $O(q_\perp^2)$
 is rather simple:
 \be
L_{11}= {\sin^2\theta \over 2}[ k_t^2- q_t^2 \cos^2(\psi-\phi) +O(q_t^4/k_t^2)]
\ee
 
The MSL rate is proportional to the spectral density of the stress tensor, integrated over the phonon momentum. We will first discuss the properties of the spectral function, and eventually compare the MSL rate with the quark annihilation process.

\subsection{Sounds} \label{sec_sounds}

   Matter produced in heavy ion collisions is not in global equilibrium; even if locally equilibrated, it expands hydrodynamically. 
In individual events, there are local density perturbations, due to quantum fluctuations in the locations of the individual nucleons.
Propagation of small perturbations on top of the expanding fireball has been studied e.g. on top of the so-called Gubser flow
\cite{Gubser:2010ui}, the corresponding Green function has been evaluated in \cite{Staig:2011wj}. The correlator of the two stress tensors
can be expressed, in the Langevin form of hydrodynamics \cite{Kapusta:2011gt}, through the two Green functions connecting
two observation points to the origin of the signal,  i.e. as the Langevin noise term. Since lower harmonics of sounds (small $k$)
have very small dissipation, given the time available $\tau_{lifetime}\sim 10 \, fm$, they can travel far across the fireball.
Thus the correlator is very nonlocal, and quite complicated, compared to the correlator in matter at rest
that we discussed in the preceding subsection.
   Fortunately, such complicated expressions for the sounds generated in flowing matter by thermal fluctuations
are not really needed. Phenomenology of heavy ion collisions is yet to discover this phenomenon. What is actually 
 observed, which is in agreement with theoretical expectations such as \cite{Staig:2011wj}, 
is so-to-say a ``boundary term", the sounds induced by the fluctuating initial conditions at the  initiation surface.
\vskip0.3cm

\begin{figure}[t]
  \begin{center}
  \includegraphics[width=8cm]{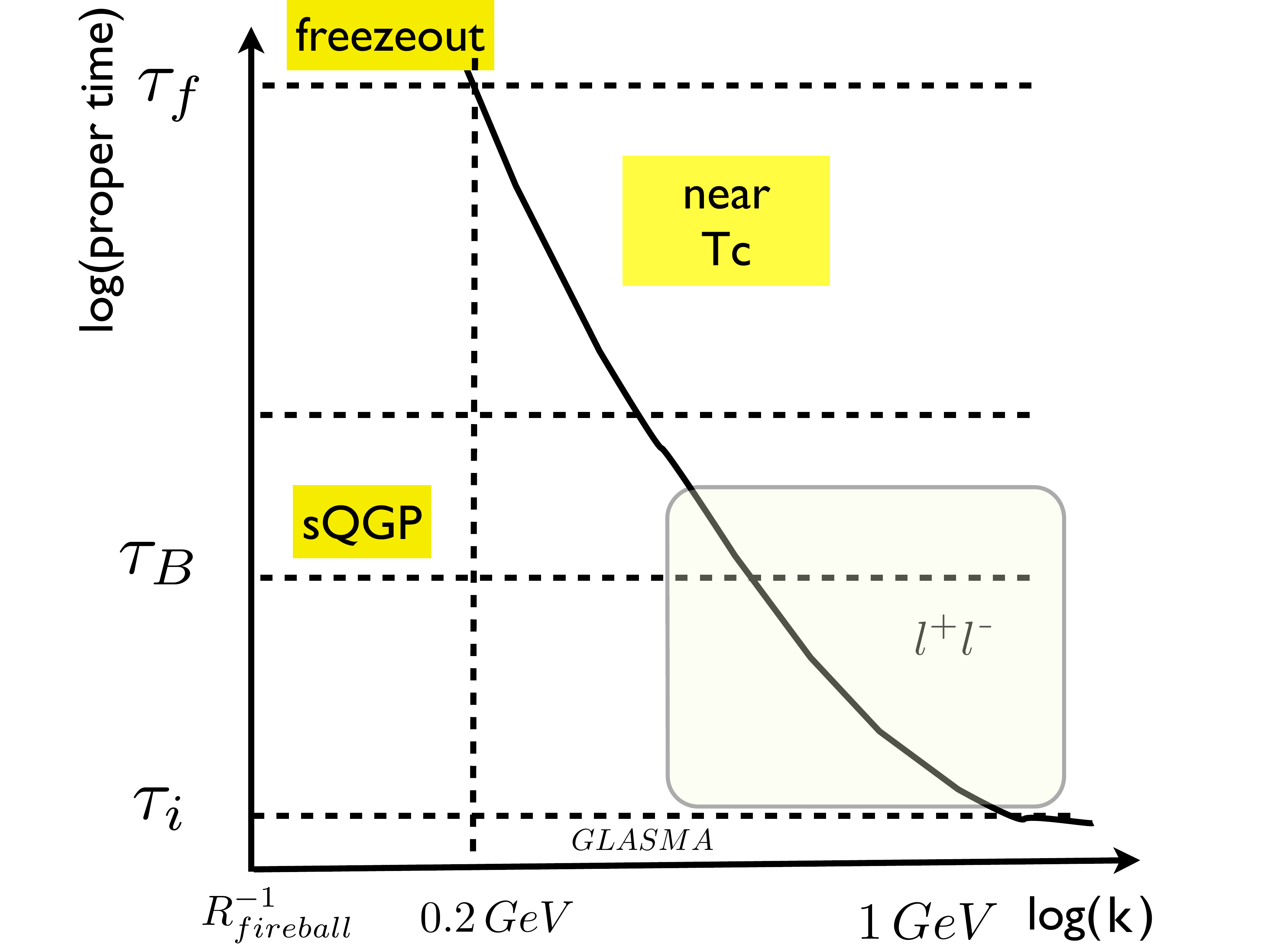}
  \caption{ (Color online)Schematic map of sounds, on a log-log plane of proper time versus the transverse momentum. For description, see text.  }
  \label{fig_sounds}
  \end{center}
\end{figure}

     The schematic map of the sound perturbations is shown in Fig. \ref{fig_sounds}. The horizontal axis is (the log of)  the perturbation momentum: it ranges from the smallest momentum, the inverse of the fireball dimensions $R_{fireball}^{-1}\sim 30\ {\rm MeV}$
   to the highest one $\sim 1 \, {\rm GeV}$.  The vertical axes is (proper) time, starting from the collision moment and ranging up to the final freeze-out of secondaries at $\tau_f\sim 10 \, fm/c$.  The most important element of this map is the solid line
   which shows how viscous damping (by a factor $1/e$) depends on the phonon momentum. It is based
   on acoustic damping in QGP (for theory see e.g.  \cite{Staig:2010by} and for phenomenology \cite{Lacey:2013is})
\be 
\left|{ \delta T_{\mu\nu} (k,t) \over  \delta T_{\mu\nu} (k,0)}\right|= \exp\left(-{2 \over 3} {\eta \over s} {k^2 t \over  T }\right) , 
\label{eqn_visc_filter}
\ee 
which can be rewritten as $\exp(-t/t_{damp})$, with the characteristic damping time 
\be  t_{damp} \equiv {3 \over 2} {s\over \eta} {1 \over k^2 T}.
\ee
This is  the solid curve plotted in  Fig. \ref{fig_sounds}. 
\vskip0.3cm

At small $k$ this time is large, and can be longer than the freeze-out time $\tau_f$. In general, in the region $below$ this curve
$t<t_{damp}$, such sound waves
 do survive viscous damping, from the beginning to the freezeout time. Therefore, in this region the stress tensor fluctuations
 are dominated by the initial fluctuations, as we discuss below.
\vskip0.3cm

 In the region well  $above$ this curve $t>t_{damping}$, the initial perturbations are already damped. The
stress tensor perturbations in this region are described by equilibrium fluctuations with the stress tensor spectral densities; for a review, see e.g. \cite{Meyer:2011gj}.
 \vskip0.3cm
 
In Fig. \ref{fig_sounds} we indicate the kinematical regions in which we will discuss the 
dilepton and photon production. The relevance of the scale $k\sim 200 \, {\rm MeV}$ follows from the fact that it corresponds to  
 the largest observed flow perturbations. Indeed, the angular harmonics with $n_{max}=6$ have 
the smallest wavelength of  those sounds, which is $2\pi R_{fireball}/n_{max}\sim 6 \, fm$, or equivalently $k\sim 200 \, {\rm MeV}$.
 The studies of the fluctuations have been done in connection to higher harmonics of the flow \cite{Alver:2010gr,Staig:2011wj}, which are also well described hydrodynamically.
 Recent work \cite{Gale:2012in} shows that model distribution over initial values of the angular moments $P(\epsilon_n)$, combined with hydrodynamics, also
 describe well the distributions of the observed flow components $P(v_n)$ for n=2,3,4. The normalized distributions have nearly a universal form  $P(\epsilon_n/<\epsilon_n>)$,
 the same for all $n$, which is rather wide, with a non-Gaussian tail toward the larger values. 

\vskip0.3cm

The sounds with higher momenta $k > 200 \, {\rm MeV}$ 
have not yet been observed. But of course they still
can produce dileptons of interest.  The proof of their existence may become an important discovery.
But how large a momentum of a phonon can be? At which point does the notion of the sound modes lose its meaning?
\vskip0.3cm

One historically important observation comes from the theory of solids; as observed by Debye, 
the cutoff $k_{max}$ should be determined by the condition that the total number of degrees of freedom in phonons should not exceed the total number of particles
\be 
V \int^{k_{max}} {d^3 k  \over (2\pi)^3}\leq \langle N_{particles}\rangle 
\ee
In other words, the sound wavelength cannot be shorter than the interparticle distance.
Applying this idea to the ``hot glue", one would argue that $k_{max} < T N_c^{2/3}$ where $N_c$ is the number of colors.
But perhaps it is incorrect to mix colorless and colored degrees of freedom. 
Indeed, the gauge theories in the limit of large number of colors, $N_c\rightarrow \infty$, such as the ones described by the AdS/CFT correspondence, yield an expression for  $k_{max}$ that does not involve $N_c$. While the interparticle distance goes to zero in this limit, the 
``end of the sound" is dictated by the imaginary part of the dispersion relation becoming comparable to the real part.
The AdS/CFT correspondence tells us that this actually happens at $ k_{max}\sim \pi T $.
\vskip0.3cm

Whether  the phonon is or is not a well-defined quasiparticle, its occupation numbers at high frequencies are in any case limited at least by the thermal weights
\be f(k^0)\approx \exp(-k^0/T)=\exp(-c_s k /T), \ee 
providing a standard thermal cutoff at high energies. So, the upper
momentum of the phonons is given by  $k < T/c_s$.   
 
\vskip0.3cm
Since there are two pictures of the initial state, the ``perturbative hot glue" and the ``glasma", let us discuss them  step-by-step, starting with the former case. The
 frequency integral for the rate will get the dominant contribution from the sound peak in the spectral function (we ignore the divergence at large $\omega$ for now). 
The width of the peak is $\delta \omega \sim \Gamma_s k^2$ and the height is
$\sim 1/\delta \omega$, so it is clear that the integral is independent of the width. Indeed, by changing $\omega\rightarrow c_s k$
anywhere except the denominator, we get a Breit-Wigner shape which is easily integrated over, yielding a simple answer
\be \int_{peak}d\omega {\rho_L \over \omega}={\epsilon+p \over 2}. \ee

The remaining integral over $d^3k$ leads us to the UV issues that are related to the ``end of sound" which we discussed above. All the previous formulae in this section were derived within hydrodynamics, assuming small $\omega,k \ll T$. 
When this assumption is no longer true,  not only  the sound is modified to a highly dissipative and thus a rapidly equilibrating mode, but also its amplitude is relaxed to that of the thermal excitations. Therefore the correlator gets its Botzmann factor $\exp(-\omega/T)$ which provides a standard thermal cutoff. Noting that the remaining integral is nothing but Debye's phonon thermal energy
\be \int {d^3k \over (2\pi)^3} k e^{-c_s k/T}= {3T^4 \over c_s^4\pi^2}, \ee
we get the following expression for the integrated sound correlator
\be \int {d^4k \over (2\pi)^4} <\epsilon_k \epsilon_k>_{corr}= (\epsilon+p) {3T^4 \over c_s^4 4\pi^3} \approx {3 T^8N_c^2 \over 5\pi},
\ee
where in the last expression we used the equation of state of ideal gluon gas and ignored the $1/N_c^2$ correction.

\subsection{Stress tensor correlators in equilibrium} 
The fluctuation-dissipation theorem, which originates from Einstein's famous Brownian motion paper,
states that in equilibrium, the dissipation rate of fluctuations should be balanced by the production rate of new ones.
And it is the latter, the source of fluctuations, that we need, as it enters
in any production process, including absorption of the ``phonons" in the media, and their conversion to light in an external magnetic field.

Hydrodynamics governs  the low energy excitations. Two spectral densities, the longitudinal (or the sound one), and the transverse
(the diffusive one) are usually defined as the basic ones. 
 For a general discussion and definitions of all stress tensor correlators see e.g. \cite{Teaney:2006nc} and --
from a lattice perspective -- \cite{Meyer:2011gj}. The imaginary part of the retarded correlator divided by $\pi$ is  known as  the spectral density. 
The equilibrium values for its longitudinal and transverse parts are
\be \rho_{L,eq}(\omega,\bk)= {(\epsilon+p) \over \pi} {\omega^3 \Gamma_s k^2 \over  (\omega^2-c_s^2 \bk^2)^2+(\omega \Gamma_s \bk^2)^2}\ee
and 
\be \rho_{T,eq}(\omega,\bk)= {1 \over \pi} { \eta\omega k^2 \over \omega^2+(\eta \bk^2/(\epsilon+p))^2},\ee
where  \be \Gamma_s={4\eta/3+\zeta \over \epsilon+p} \ee is the sound attenuation length.
The correlator of the energy densities $\epsilon=T^{00}$ is related to the spectral density via
\be \chi_{\epsilon\epsilon} ={\omega^2 \over k^2}  \rho_{L,eq}(\omega,\bk) \ee
Note that $k^2$ in the numerator then cancels, which allows to get $k\rightarrow 0$ limit first, and $\omega\rightarrow 0$
second, as required for the Kubo formula.
\vskip0.3cm

The trace of the stress tensor $\theta=T^\mu_\mu$ is a special combination, related to trace anomaly and discussed in \cite{Basar:2012bp},
its spectral density is
\be
 \rho_{\theta}= {9 \omega \over \pi}  \left( \zeta+ (1/3-c_s^2)^2{(\epsilon+p) \,\Gamma_s  \bk^4 \over  (\omega^2-c_s^2 \bk^2)^2+(\omega \Gamma_s \bk^2)^2}\right).
\ee
At $T \gg T_c$, the sQGP is approximately conformal, and thus this combination is strongly suppressed.
The small factor $(1/3-c_s^2)^2\sim 10^{-2}$ is the consequence of this fact. The same small factor is actually present in the first term -- the bulk viscosity $\zeta$, as well.
Therefore,  this contribution is in fact small compared to that of the $L$ and $T$ spectral densities, except for the region $T \sim T_c$, where the bulk viscosity is expected to be relatively large \cite{Kharzeev:2007wb,Karsch:2007jc,Meyer:2007dy,Romatschke:2009ng}.

 \subsection{Stress tensor out of equilibrium}
    We need the momenta of stress tensor perturbations to be as large as possible, limited only by our assumption $q_t \ll M$. 
 Such ``ultrasounds" have a short damping time by viscosity which prevents them from being observed
  at freezout, in correlations of the secondary hadrons. Nevertheless, there is no doubt that the quantum fluctuations inside the colliding nuclei/nucleons do produce these sounds at the collision time, and for the early time dilepton production processes, especially for the MSL process driven by the short-lived magnetic field,  they are important.
  Without going into a specific modeling, we can argue that at the momentum scale of interest $k\sim 1\, {\rm GeV}$,
  the early time stress tensor perturbations are of order one
  \be {\delta T^{\mu\nu} \over  <T^{\mu\nu} >} \sim 1. \ee
  This  means that the total yield of the dileptons due to the MSL and MTL processes should be comparable.
  Parametrically, the ratio is only slightly suppressed by the ratio of the number of colorless ``hydrodynamical modes"
  in the correlation function to the total number of
  gluonic modes (in the energy density),
  \be {MSL \over MTL} \sim {3 \over 8}. \ee
 united into a stress tensor.

 \section{Summary and Discussion}
 \label{sec_conc}
   The stress tensor and its perturbations -- Lorentz spin-2 tensors -- can be connected (via the quark virtual loop)
   to two photons, one of which can be a coherent magnetic field present at non-central heavy ion collisions. 
Such processes, if identified and studied, can be used as valuable diagnostic tools of produced matter.
\vskip0.3cm

The terminology proposed in this paper identifies two distinct processes:
{\em Magneto-Thermo-Luminescence} (an interaction of  $\vec{B}(x,t)$ with near-constant 
average stress tensor $\langle T_{\mu\nu}\rangle$, and
{\em Magneto-Sono-Luminescence}  
(an interaction of $\vec{B}$ with perturbations of the stress tensor $\delta T_{\mu\nu}(k)$ possessing a certain 4-momentum $k$).
\vskip0.3cm

We argued that the most promising application of these phenomena is the contribution to the 
``intermediate mass dileptons" (IMD)  (defined in (\ref{IMD})). 
We calculated the rates of these processes, in some simplified kinematics, and indicated their qualitative features which can be used
to separate them from other production mechanisms.
The magnitude of the process calculated by the hadronic approach -- as shown in Fig.\ref{fig_MTL_qq} -- is below the magnitude of the lowest order process roughly by an order of magnitude. The  OPE approach, on the other hand, leads to a rate which is comparable to the lowest order quark annihilation. However
 these estimates depend on many factors, including e.g. the unknown quark fugacity, and provide  qualitative guidance only. 
\vskip0.3cm

How can the contribution of the MSL process be experimentally separated from others?
The first distinctive feature of MSL is a very specific {\em centrality} dependence. While most other processes
are maximal at central collisions, the MSL is zero at $b=0$ and peaks at the impact parameter
value $b\sim 10 \, fm$ . Of course, this is because the yield of MSL is proportional to the coherent magnetic field (one should of course keep in mind also the role of fluctuations).
\vskip0.3cm

   Another rather  unusual property of these processes is that they produce virtual photons which are {\em longitudinally} polarized, since the dilepton angular distributions are proportional to $\sin^2\theta$. In terms of the polarization parameter $a$ defined via $(1+a \cos^2\theta)$,
the produced dileptons are characterized by $a=-1$. 
    For comparison, let us remind the reader what is the angular dependence for other dilepton production processes.
 Classic partonic Drell-Yan process at leading order yields $a=1$, corresponding to transverse photon. Thermally equilibrated medium would produce unpolarized photons and thus no 
 theta dependence, $a=0$ . It has been recently argued by one of us  \cite{Shuryak:2012nf},  that the lowest order $\bar{q}q$ processes in
 a  {\em non-equilibrium} plasma can also lead  to a (smaller in magnitude) negative $a$ in some interval of invariant mass, albeit with no anisotropy with respect to the heavy ion reaction plane. 
 \vskip0.3cm
 We urge further experimental studies of the angular dependence of photon and dilepton production at RHIC and LHC, and hope that they would allow to establish the existence of magnetosonoluminescence in quark-gluon plasma. 
 
 \vskip0.3cm

{\bf Acknowledgements}
This work was supported in part by the U.S. Department of Energy under Contracts No. DE-FG-88ER40388 and DE-AC02-98CH10886.

 \appendix
 
\section{Notations} In this paper we use the same normalizations for both QED and QCD couplings and fields, so  that $e^2/4\pi=\alpha\approx 1/137$, $g^2/4\pi=\alpha_s$
and e.g. the energy density of both gluonic and electromagnetic  fields is given by $T^{00}= (1/2) (\vec{E}^2+\vec{B}^2)$.

\section{Random Gaussian glue}
\label{app_glue}
 
One  may assume a Gaussian ensemble of the gluonic fields, in which only $colorless$ quadratic combinations of the fields are nonzero. 
 We will also make average over orientations of the fields in the amplitude, denoted by angular brackets.
 In 3-dimensional notations,
 in the rest frame of matter one can think of three scalars, $\vec{E}^2,   \vec{B}^2,\vec{E}.\vec{B}$. In 4-dimensional notations, the corresponding 3 structures 
 can define the following decomposition of the field strength: 
 
\ba <{G}^m_{\mu_1 \mu_2} {G}^n_{\mu_3 \mu_4}>&=&\delta^{mn}[A P^a_{\mu_1 \mu_2 \mu_3 \mu_4} +B P^b_{\mu_1 \mu_2 \mu_3 \mu_4}  \nonumber \\
&&\qquad+C P^c_{\mu_1 \mu_2 \mu_3 \mu_4}] \nonumber \\
    P^a_{\mu_1 \mu_2 \mu_3 \mu_4}&=&\epsilon_{\mu_1 \mu_2 \mu_3 \mu_4}   \nonumber \\
    P^b_{\mu_1 \mu_2 \mu_3 \mu_4}&=&g_{\mu_1,\mu_3} g_{\mu_2,\mu_4}-g_{\mu_1,\mu_4}g_{\mu_2,\mu_3}  \nonumber \\
 P^c_{\mu_1 \mu_2 \mu_3 \mu_4}&=&n_{\mu_1} n_{\mu_3} g_{\mu_2,\mu_4} +n_{\mu_2}n_{\mu_4}g_{\mu_1,\mu_3}  \nonumber \\
    &&-n_{\mu_1}n_{\mu_4}g_{\mu_2,\mu_3}-n_{\mu_2}n_{\mu_3}g_{\mu_1,\mu_4}   \nonumber \\
    &&+(1/2)   P^b_{\mu_1 \mu_2 \mu_3 \mu_4}. \label{eqn_GG}
 \ea
 These three tensors are orthogonal to each other.  The meaning of the three structures follows from convolutions of these expressions with $P^a,P^b, P^c $, respectively:
 \be <{G}^m_{\mu_1 \mu_2} {G}^n_{\mu_3 \mu_4}> P^a_{\mu_1 \mu_2 \mu_3 \mu_4}=-24A, \ee 
  \be <{G}^m_{\mu_1 \mu_2} {G}^n_{\mu_3 \mu_4}> P^b_{\mu_1 \mu_2 \mu_3 \mu_4}=24 B, \ee 
  \be <{G}^m_{\mu_1 \mu_2} {G}^n_{\mu_3 \mu_4}> P^c_{\mu_1 \mu_2 \mu_3 \mu_4}= 6 C, \ee
 which defines
three parameters $A,B,C$ that, in Gaussian approximation,  contain the entire information about the local properties of the ``Glasma".  
 
\vskip0.3cm

The $dual$ field strength is defined as 
 \be \tilde{G}_{\mu\nu}=(1/2)\epsilon_{\mu \nu \alpha \beta} G_{\alpha \beta}. \ee
The action of the electric-magnetic duality transformation is more clear in the usual 3-d notations, in which it is simply an interchange of electric and magnetic fields. 
Under it, the combinations
 \ba A &\sim& \vec{E}\cdot\vec{B} \\
B&=&G^a_{\mu\nu}G^{a,\mu\nu}{1 \over 12( N_c^2-1)} \\
 C&=& {2 \over 3 (N_c^2-1)} T^{00}= {1 \over 3(N_c^2-1)} (\vec E^2 +\vec B^2) \ea 
 remain unchanged, while the second one $B\sim (\vec{E}^2 -   \vec{B}^2)$ obviously changes sign. (In this expressions we have taken matter at rest.)
Therefore one finds 
\ba <\tilde{G}^m_{\mu_1 \mu_2} \tilde{G}^n_{\mu_3 \mu_4}>&=&\delta^{mn}[A P^a_{\mu_1 \mu_2 \mu_3 \mu_4}-B P^b_{\mu_1 \mu_2 \mu_3 \mu_4} \nonumber\\
&&\qquad +C P^c_{\mu_1 \mu_2 \mu_3 \mu_4}] .\ea

\section{ Calculating components of the self energy}

Acting, as usual, by the tensor structures $P_1, P_2$ on the polarization tensor one finds that (since $P_1$ and $P_2$ are not mutually orthogonal) 
a system of two linear equations
\ba
\Pi_1=\Pi_{\mu\nu} P^1_{\mu\nu} =P^{11} W_1+P^{12} W_2\nonumber \\
\Pi_2=\Pi_{\mu\nu} P^2_{\mu\nu} =P^{12} W_1+P^{22} W_2 ,
\ea
where (for simplicity) we give the coefficients in the rest frame of the matter, $n_0=1, n_1=n_2=n_3=0$
\ba P^{11} =P^1_{\mu\nu} P^1_{\mu\nu} =3 \\ \nonumber
 P^{22} =P^2_{\mu\nu} P^2_{\mu\nu} =  {{\bf q}^4 \over (q_0^2-{\bf q}^2)^2 }\\ \nonumber
 P^{12} =P^1_{\mu\nu} P^2_{\mu\nu} =-{{\bf q}^2 \over q_0^2-{\bf q}^2 }.  \ea
The solution is obvious, we only note that in order for it to exist the determinant of the system
\be  P^{11}  P^{22} -P^{12}P^{12}=2{{\bf q}^4 \over (q_0^2-{\bf q}^2)^2 } \neq 0 \ee
should be nonzero, thus the spatial part of the vector $q$ cannot vanish.

For the color averaged gauge fields, the result is 
 \ba  \delta\Pi_1&=& c_{1B} B  + c_{1C} C \\ \nonumber
   \delta\Pi_2&=& c_{2B} B  + c_{2C} C \\ \nonumber
   c_{1B}&=&- 3(q_0^2-{\bf q}^2) \\ \nonumber
      c_{1C}&=&1/2 (3 q_0^2+{\bf q}^2)\\ \nonumber
         c_{2B}&=& {\bf q}^2 \\ \nonumber
            c_{2C}&=&-(1/2) {{\bf q}^2 (q_0^2+3 {\bf q}^2)  \over q_0^2-{\bf q}^2} . \nonumber
 \ea
 Solving this $2\times 2$ system, we get our effective action
 \be L_{eff}= \Pi_{\mu\nu} A_\mu A_\nu. \ee

\section{Fourier transforms}

In order to evaluate the Fourier transforms we use the standard dimensional regularization scheme where $d=4+2 \epsilon$. Another useful tool is the proper time representation:
\be
\frac{1}{(q^2)^{n+1}}=\frac{1}{\Gamma(1+n)}\int_0^\infty ds s^n e^{-q^2 s}
\ee
By standard integrations we obtain
\ba
&&\frac{q_\mu}{q^4}\rightarrow\frac{i x_\mu}{8 \pi^2 x^2}\quad,\quad \frac{q_\mu}{q^6}\rightarrow-\frac{i x_\mu}{64 \pi^2}\left(\ln x^2-\frac{1}{\epsilon}\right) \nonumber \\
&&\frac{q_\mu q_\nu q_\lambda}{q^8}\rightarrow-\frac{i}{192 \pi^2}\left[\frac{x_\mu x_\nu x_\lambda}{x^2} \right.\nonumber \\&&\left.+\frac{1}{2}\left(x_\mu g_{\nu\lambda}+x_\nu g_{\mu\lambda}+x_\lambda g_{\mu \nu}\right)\left(\ln x^2-\frac{1}{\epsilon}\right) \right].
\ea
The relevant transforms to convert the self energy (\ref{selfx}) from position to momentum space are:

\ba
&&\frac{x_\mu x_\nu}{x^4}\rightarrow\frac{2\pi^2}{q^4}(g_{\mu\nu}-2q_\mu q_\nu)\nonumber\\
&&\frac{x_\mu x_\nu x_\alpha x_\beta}{x^6}\rightarrow\frac{\pi^2}{2\,p^6}\left[8 q_\mu q_\nu q_\alpha q_\beta\right.\nonumber\\
&&-2q^2(g_{\mu\nu}q_\alpha q_\beta+g_{\mu\alpha}q_\nu q_\beta+g_{\mu\beta}q_\nu q_\alpha\nonumber\\
&&\hskip0.8cm+g_{\alpha\beta}q_\mu q_\nu+g_{\nu\beta}q_\mu q_\alpha+g_{\nu\alpha}q_\mu q_\beta)\nonumber\\
&&+q^4(g_{\mu\nu}g_{\alpha\beta}+g_{\mu\alpha}g_{\nu\beta}+g_{\mu\beta}g_{\nu\alpha})\nonumber\\
&&\frac{x_\mu x_\nu}{x^4}\ln x^2\rightarrow\frac{2\pi^2}{q^4}[(2q_\mu q_\nu-q^2g_{\mu\nu})\ln (q^2/4)\nonumber\\&&-4q_\mu q\nu+g_{\mu\nu}q^2]
\ea

 \section{The lowest order rate}
 Even though the total and differenttial rates in the lowest (zeroth) order in strong interactions coupling, $\alpha_s$, has been known for a long time \cite{Shuryak:1978ij},
 it is useful to put it in a form similar to  the expressions in this paper, so that the common factors drop in the ratio.
 Diagrammatically the calculation consists of a convolution of quark and lepton ``cut loops" 
 shown in Fig.\ref{fig_diag}(a)
 which corresponds to the following expression
 \ba
  {dN \over d^4x}  &=& e^4\sum_f Q_f^2 N_c
 {Q_{\mu\nu} L_{\mu\nu} \over M^4} 
  f_F({q_{+0} \over T})  f_F({q_{-0} \over T})\nonumber\\
  &&\times PS(l^+,l^-) PS( q^+,q^-) \nonumber\\
  &&\times(2\pi)^4 \delta^4(q_++q_-- l_+-l_-)
   \ea
   where 
\ba PS( q^+,q^-) = {d^3 q_+ \over (2\pi)^3  q_{+0} } {d^3 q_- \over (2\pi)^3  q_{-0} },\ea
 and the same for leptons. The  fermion loop summed over polarizations for massless leptons 
is
 \ba
 L_{\mu\nu}&=& (1/4)Tr\left( \gamma_\mu \gamma_\alpha \gamma_\nu \gamma_\beta   \right) l^+_\alpha l^-_\beta \nonumber\\
  &=&l^+_\mu l^-_\nu+l^+_\nu l^-_\mu-g_{\mu\nu} (l^+l^-).\label{Lmunu}
 \ea
$Q_{\mu\nu}$ is the  same tensor for quarks, with appropriate change of momenta. The covariant convolution $Q_{\mu\nu} L^ {\mu\nu} $ expressed in the center of mass frame of the dilepton
is $M^4(1+\cos^2\theta)$, where $\theta$ is the angle between the quark and lepton directions in this frame.
For isotropic quark distribution the latter factor averages to $4/3$.  

 The Fermi distribution functions $f$ stand for distributions of the original quark and antiquark. For arbitrary kinematics of the dileptons the integral over quark phase space is complicated.
  However, in the dominant kinematics we use for comparison -- the dilepton as a whole is at rest in matter rest frame -- this integration becomes trivial. 
     In Boltzmann approximation the thermal distribution reduces  to \be f_F(\epsilon^+/T)f_F(\epsilon^-/T) \approx \exp(-{M\over T}).\ee

\end{document}